\documentclass[conference]{IEEEtran}

\usepackage{amsmath,amssymb,mathtools}
\usepackage{array,booktabs,multirow,tabularx}
\usepackage{cite}
\usepackage{graphicx}
\usepackage{xcolor}
\usepackage{url}
\usepackage{microtype}
\usepackage[hidelinks]{hyperref}

\graphicspath{{figures/}}
\newcommand{\systemname}{\textsc{KeyPooling}}

\newcommand{\evidence}[1]{\textsc{#1}}
\newcommand{\mzero}{\textnormal{\textsc{M0}}}
\newcommand{\aone}{\textnormal{\textsc{A1}}}
\newcommand{\atwo}{\textnormal{\textsc{A2}}}

\newcolumntype{Y}{>{\raggedright\arraybackslash}X}
\newcolumntype{L}[1]{>{\raggedright\arraybackslash\hsize=#1\hsize}X}
\newcolumntype{C}[1]{>{\centering\arraybackslash\hsize=#1\hsize}X}
\newcolumntype{N}[1]{>{\raggedleft\arraybackslash\hsize=#1\hsize}X}

\hypersetup{
  pdftitle={KeyPooling: Measuring Where LLM API Relay Paths Collapse Prompt Cache Isolation},
  pdfauthor={Bowen Sun; Yixi Cai; Xiaogeng Liu; Zhengyue Zhao; Yinzhi Cao; Chaowei Xiao},
  pdfsubject={Measurement of prompt cache isolation in LLM API relay paths},
  pdfkeywords={LLM API relays, prompt caching, cache isolation, security measurement}
}

\begin{document}

\title{KeyPooling: Measuring Where LLM API Relay Paths Collapse Prompt Cache Isolation}

\author{
\IEEEauthorblockN{
Bowen Sun\textsuperscript{1,*},
Yixi Cai\textsuperscript{2},
Xiaogeng Liu\textsuperscript{1},\\
Zhengyue Zhao\textsuperscript{1},
Yinzhi Cao\textsuperscript{1},
and Chaowei Xiao\textsuperscript{1}}
\IEEEauthorblockA{\textsuperscript{1}Johns Hopkins University\\
\textsuperscript{2}The Chinese University of Hong Kong\\
Email: \texttt{bsun39@jh.edu}\qquad \textsuperscript{*}Corresponding author}
}

\maketitle

\begin{abstract}
Large language model (LLM) API relays authenticate customers separately but often forward requests through shared provider credentials. Providers scope prompt caches to upstream principals and namespaces, so relay customers mapped to one cache identity can observe each other's cache state. Prior work showed cache sharing at selected endpoints but did not identify which credential, pool, adapter, or nested hop controls the final identity. We present KeyPooling, a measurement method that traces customer identity through cache lookup and write, verifies runtime transformations, and tests one predicted identity component at a time. Across five open-source gateways connected to OpenAI and Anthropic, none bound customers to upstream credentials by default; under a shared credential, all five exposed cross-customer cache reads for both providers. Principal and namespace splits, pool associations, and adapter and nested-relay contrasts localized the controlling transformations. In an outcome-independent weekly OpenRouter frame, tests covered 80.5\% of eligible token volume and found cross-account reads for 12 of 28 labels carrying 33.7\% of volume. On one production route, a controlled procedure recovered eight consecutive target positions without target access. Broader tests identify cache granularity, routing, rate limits, attribution, and budget as conditions for token-by-token recovery, not security controls. We derive a defense contract: every customer must enter a provider-enforced domain, or a namespace derived from authenticated identity must survive every final cache lookup and write. Placing this split after reusable public prefixes preserved most modeled reuse at a 1.7--2.5\% cost increase.

\end{abstract}

\section{Introduction}
\label{sec:introduction}

Large language model (LLM) API relays authenticate customers, meter usage, translate request schemas, choose models and providers, and pay the upstream bill; throughout, \emph{upstream} means toward the provider and \emph{downstream} toward the customer.  A relay key therefore appears to identify a domain the provider enforces.  The model provider may see many relay customers through one credential, project, workspace, route, or namespace.  Prompt cache lookup uses the identities visible at the provider boundary.  Customer separation therefore depends on whether the relay preserves the original identity through that final boundary.

\begin{figure*}[t]
  \centering
  \includegraphics[width=\textwidth]{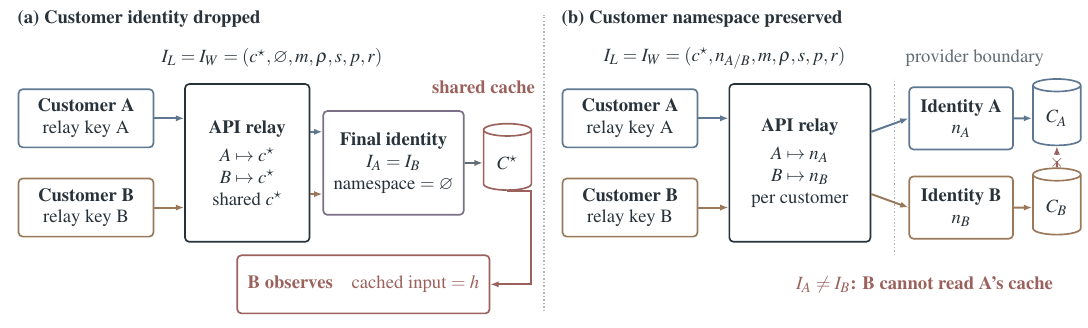}
\caption{Downstream authentication and final cache identity can diverge.  Pooling maps customers A and B into one cache domain (left), while a namespace the relay derives from the original customer keeps them apart and preserves reuse by the same customer (right).}
  \label{fig:identity-collapse}
\end{figure*}

Consider two mutually distrustful customers, Alice and Bob.  Alice submits a prompt $P$ long enough to be cached, and the provider creates reusable cache state.  When Bob later submits the same $P$ through a different relay account, a positive cached input count tells him that someone already sent $P$, which reveals a fact about Alice's prompt history such as whether a confidential template is in production use.  If Bob knows a prefix $H$ but not what follows it, he can test extensions $Hc_1,Hc_2,\ldots$, append whichever one the feedback identifies, and repeat.  We call the loss of original customer identity at the final cache boundary \emph{key pooling}, and name our method \systemname{} after it.  Key pooling leaks membership for known prefixes, and it supports sequential recovery wherever the path gives precise and stable feedback.  Figure~\ref{fig:identity-collapse} contrasts a pooled path with one that carries the customer's identity through to the provider.

Prompt cache admission usually requires a long matching prefix, and public agent harnesses supply one because system instructions, tool descriptions, and schemas precede the private request.  All 17 public Claude Code, Cursor, and OpenCode prefixes in our frozen corpus exceed 1,024 tokens under the Grok-2 tokenizer, so current agent requests clear common admission floors before private content begins.

Relay adoption makes the boundary relevant at production scale.  OpenRouter reports more than 200 trillion tokens per month and 10 million users, and an independent model places its earlier run rate near 2\% of global monthly LLM token consumption~\cite{openrouter_about,borri2026aipremium}.  Ramp finds OpenRouter among roughly half of category purchasers, and our reproducible search retained 29 open source relays and gateways~\cite{ramp_openrouter}.

Prior work establishes the component attacks.  Early Bird and PROMPTPEEK reconstruct cached prefixes in controlled serving environments, InputSnatch and OptiLeak improve candidate construction, and Gu et al. audit cache sharing through production provider APIs~\cite{song2025earlybird,wu2025promptpeek,zheng2024inputsnatch,wang2026optileak,gu2025auditing}. CacheProbe observes reuse across OpenRouter accounts on managed routes and separation when each customer supplies an independently scoped provider key~\cite{fahey2026cacheprobe}.  Those two endpoints differ in principal ownership, capacity, scheduling, route, and affinity at the same time, so the contrast shows that exposure exists without showing which transformation causes it.

Our central insight is that prompt cache isolation is a property of a concrete final path rather than of a relay account, project, or model label.  \systemname{} audits a path in three stages.  Source tracing follows the authenticated customer through credentials, namespaces, adapters, pools, retries, fallbacks, and nested relays.  A runtime observer verifies the identity and cache controls that leave the gateway.  A controlled cache test then changes one predicted identity component while holding the prompt, schedule, and model label fixed.  In our measurements, observations were determined by the selected pool partition, and an inner overwrite reopened access that an outer namespace had removed.

Three ambiguities make those controls necessary: a response with no reported reuse may mean isolation or merely a failed lookup, a managed endpoint moves several platform components at once, and an extraction probe writes its own candidate into the cache.  Section~\ref{sec:methodology} resolves each in turn.

Our contributions are:

\begin{itemize}
\item We formulate relay cache confidentiality as preservation of authenticated customer identity through every final lookup and write.  Runtime verification and matched interventions localize the path that loses identity.
\item We measure the mechanism through six frozen source revisions, five of them executable, ten experiments with real upstream models, a ranked OpenRouter frame, and production deployments stratified by framework.  No gateway binds a customer to an upstream credential by default, independently operated deployments reproduced the effect in two of the five executable projects, and real Gemini creation requests show that exposure also needs the relay to delegate the provider's cache lifecycle.
\item We establish a measured limit from membership of a known prefix to sequential recovery.  Primary and shadow histories account for attacker cache writes, and one production route supported recovery of a controlled target carrying 16 bits.
\item We derive a deployment obligation for every final path, validate the routing and failure behavior it requires on all five executable gateways against a local fixture, and prototype the relay derived namespace on one gateway against a real upstream.  A split after reusable public prompt regions separates private suffix state across 5,713 replayed request events while preserving most shared prefix reuse.
\end{itemize}


\section{Motivating Example and Threat Model}
\label{sec:background-threat}
\label{sec:motivating-example}

This section explains how a shared provider cache becomes an observable signal between relay customers and states the exact attacker model used throughout the paper.  We first build the attack from a concrete four request example.  We then show why attribution requires the complete path, trace the relay operations and provider interfaces that control cache state, and conclude with the attacker's capabilities, objectives, and experimental scope.

\subsection{One relay, two customers, one provider cache}

Prompt caching reuses the key and value state computed for a sufficiently long prompt prefix, and creating that state is \emph{cache formation}; an application response cache instead returns a stored response.  Providers expose reuse through counts of cached input, counts of cache reads and writes, price, latency, headers, quotas, or state visible to later calls.  We primarily use the \emph{cached input count}, a documented integer visible through the API, which lets us avoid a classifier based on network timing.

Loss of customer separation becomes an information leak when one customer's response depends on another customer's private cache history.  We name three levels, because each stronger one needs conditions the previous one does not.  \mzero{} is the mechanism: two downstream customers arrive at one final cache identity.  \aone{} is the first observable capability, and it holds when cache formation, reachability, and a visible report together tell the attacker whether a known prefix $P$ is present, which separates a history in which the victim, in our experiments a second controlled customer, submitted $P$ from one in which the victim did not.  \atwo{} is the strongest, and it holds when precise reporting, attribution, stable routing, a cached entry that stays resident, candidate coverage, and budget let the attacker extend a known prefix adaptively, one token at a time, and so read content it did not start with.  We label every observation with the strongest level its evidence supports.

Cache admission looks like a practical barrier: the interfaces we study reuse state only once an eligible prefix reaches a minimum length the path chooses, so a probe must begin with enough matching material.  Matching is also exact, so only the span of the victim prefix that is identical on every run is usable, and content the harness resolves at run time, such as a date, a working directory, or environment metadata, ends that span.  In the agent corpus of Section~\ref{sec:introduction}, 15 of the 17 prefixes contain no resolved value at all, and the two that do place the first one after more than 98\% of the prefix.  Public system instructions, tool descriptions, and schemas come before both the private request and any resolved value, so each prefix still supplies from 1,559 to 8,794 reproducible tokens under the same Grok-2 tokenizer, above both the 256 and 1,024 token floors we observed.  Appendix~\ref{app:extended-methodology} records the variants, the coding rule this count depends on, and the tokenizer digest.

Return now to Alice and Bob.  They have separate relay accounts, but the relay maps both to one upstream credential.  For the four request unit, define the \emph{cached input vector} as $\mathbf{C}=[c_1,c_2,c_3,c_4]$, where $c_i$ is the documented count of tokens read from the cache in response $i$ after normalizing provider field names.  Thus $c_i=0$ reports no cache read and $c_i=h>0$ reports reuse of $h$ input tokens.

\emph{Operational cache vocabulary.}  Four terms recur throughout.  Let $R(q)$ be the normalized cached input count returned for request $q$, and let $\tau(q)$ be the protocol's predeclared \emph{reuse threshold}: a response is \emph{hot} when $R(q)\geq\tau(q)$ and \emph{cold} otherwise, and the four request unit sets $\tau(q)=1$.  An \emph{owner hot repeat} is a later identical hot request from the account that primed the prefix; it shows that reusable state formed and stayed observable during the trial.  A primary and shadow pair is \emph{jointly hot} when both responses clear their thresholds on comparable routes.  \emph{Strict positive} and \emph{controlled miss}, shortened to hit and miss in result tables, name the outcomes of one four request unit; one execution of that unit on a path is a \emph{cell}.

\begin{figure*}[t]
  \centering
  \includegraphics[width=\textwidth]{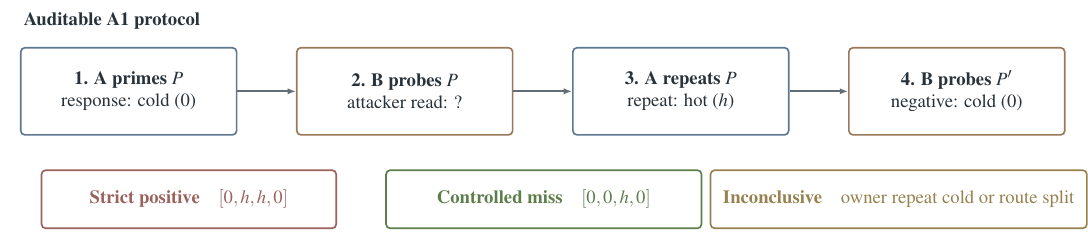}
\caption{A concrete cache observation between customers.  The second request in the sequence is the attack action.  Alice's repeat and Bob's fresh negative distinguish a strict positive from failed or indiscriminate cache formation.}
  \label{fig:a1-protocol}
\end{figure*}

Every trial uses a newly generated, long synthetic prefix $P$ with high entropy and an independently generated unrelated prefix $P'$, both above the calibrated admission floor.  Random content and freshness in each cell make an accidental match with another customer's request negligible.  In order, Alice sends $P$, Bob sends $P$, Alice repeats $P$, and Bob sends $P'$.  A strict positive between customers is $\mathbf{C}=[0,h,h,0]$: Alice's first request was cold, Bob then read state associated with $P$, Alice's owner repeat remained hot, and Bob's unrelated negative remained cold.  A controlled miss is $[0,0,h,0]$.  Figure~\ref{fig:a1-protocol} shows the four request unit.

A strict positive establishes \aone{} for that path and window.

What an \aone{} observation is worth depends entirely on which prefix $P$ the attacker holds and can reproduce, because the test itself does not care what $P$ contains.  Set $P$ to a confidential system prompt, tool schema, or agent harness, and a positive confirms that the template is in active production use.  Set $P$ to a purported leaked prompt, and a positive separates a genuine deployment from a fabrication.  Set $P$ to a specific context document, which admission accepts as readily as any other long prefix, and a positive shows that the document is processed inside the shared domain.  In every case a positive attributes $P$ to some customer sharing the final cache identity, never to a named account, and only within the residence window, the period a cached prefix survives before eviction.  Recovering a continuation that the attacker does not already hold requires the stronger \atwo{} conditions.

\subsection{Attribution requires the complete path}

The four request unit establishes an observable effect, but locating its cause means tracing the complete relay path.  A positive can come from pooled provider credentials, a namespace dropped in one adapter, the same pool partition selected twice, identity overwritten in a nested hop, or a retry onto a different route.  A zero can equally mean isolation, failed cache formation, expiry, route mismatch, or an interface that never delegated cache creation.  Each service therefore has to be characterized path by path.

Cache writes by the probe create another ambiguity.  Bob's probe can itself insert $P$, so Alice's later hot repeat cannot identify who created the entry Bob observed.  The problem sharpens during extraction, where a wrong candidate tested once may look correct on a later test simply because the attacker created reusable state for it.  The four request unit controls formation and supplies fresh negatives; the extraction method additionally compares a history containing victim state against a shadow carrying only matching attacker writes, then replays the whole schedule with no victim at all.

\subsection{Relay paths and cache authority}

The ambiguities above arise because a relay can transform identity and cache authority at several stages.  A relay terminates a downstream key, maps a model alias to a provider, selects a managed credential or a credential supplied through bring your own key (BYOK), translates schemas, and may retry or fall back.  Credential pools and nested relays repeat the choices.  One project can therefore preserve identity on a native adapter and erase identity on a generic or compatibility path.  A concrete final path is the security unit.

Providers also differ in how state forms.  Eligible OpenAI prompts can create cache state through ordinary inference.  Anthropic requires cache control in each request.  Gemini exposes implicit reuse managed by the provider and an explicit \texttt{CachedContent} object with creation, time to live (TTL), reference, and deletion semantics~\cite{openai_prompt_cache,anthropic_cache,google_gemini_cache}.  A shared upstream project therefore creates exposure only where the relay also forwards or implements the relevant cache control interface.

\subsection{Attacker, objectives, and claim scope}

The preceding path and lifecycle analysis determines what an ordinary customer can influence and observe.  Our attacker is an authorized relay or provider customer.  The attacker can submit adaptive requests with small output limits and observe ordinary usage fields, prices, timing, route metadata, errors, and future state.  The attacker may know a public harness prefix or a short correct prefix of a synthetic continuation.  The relay, upstream principal, physical shard, route, and eviction policy remain outside the attacker's control.

All experimental victims are second accounts, workspaces, teams, or downstream principals controlled by the researchers, every target is a nonfunctional synthetic canary, and active experiments query only controlled content.  We define \emph{strict isolation} as independence of every transcript, or complete observable record, visible to the attacker, including future state, from another customer's private cache history.  Under strict isolation, only a rate, route, or budget barrier enforced as a security invariant contributes to separation.

\section{Causal Path Model}
\label{sec:identity-theory}

This section explains how relay operations can preserve or erase customer separation before the provider accesses cache state.  The security unit is a concrete path to the final cache lookup and write.  We first define the identities used by those operations, then connect loss of identity to the two attack capabilities.  We next derive interventions that locate the responsible transformation and conclude with the deployment obligation that follows when many customers share fewer enforced domains.  Appendix~\ref{app:formal-results} gives supporting results on directed reachability, sequential complexity, granularity, and domain capacity.

\subsection{Final lookup and write identities}

Let $d$ be the authenticated downstream customer and $\pi$ a path through credentials, namespaces, adapters, pools, retries, fallbacks, and nested relays.  At the provider boundary, the path selects an upstream principal $c$, namespace $n$, model or deployment $m$, provider route $\rho$, affinity value $s$, pool partition $p$, and optional cache resource handle $r$.  We summarize the identities used by the final cache operations as
\begin{equation}
 I_x(\pi,d,q,H,t)=(c,n,m,\rho,s,p,r)_x,
 \qquad x\in\{L,W\}.
 \label{eq:final-identities}
\end{equation}
The arguments are inputs and the tuple is the value they select; no argument appears in the tuple.  Lookup and write can select different values within one request, and both depend on $q$, $H$, and $t$ because retries and hidden scheduling may change the selection even when the public endpoint and model label remain fixed.

The original customer is protected only if every final path either selects a genuinely separate principal enforced by the provider or carries an unforgeable customer namespace into both $I_L$ and $I_W$.  A check at the relay's entry point covers only the first transformation.  A later adapter can drop the namespace, a pool can send two customers to the same partition, and a fallback can use a weaker identity than the successful primary route. Additionally, an explicit cache object adds an availability condition: where a relay exposes neither the creation and reference operations nor an internal translation of them, the feature is simply unreachable through that path until a later adapter delegates the object lifecycle.

\subsection{From key pooling to attack capability}

Identity loss is a structural condition; we now connect it to observable attack capability.  We say that customer A \emph{reaches} customer B when a request by A can change B's later transcript through cache state.  Reaching is directed: a write on one route may become readable on another, while a retry, eviction, or route change can break the reverse direction.  A black box experiment therefore establishes a directed edge for one path and window.

Directed reachability separates the mechanism from the two attack capabilities of Section~\ref{sec:motivating-example}.  A usable sequential oracle adds five requirements to an \aone{} edge: reports that separate competing extensions instead of quantizing them together, attribution to victim state despite the attacker's own cache writes, a route and entry that survive many dependent probes, a candidate set that contains the truth, and a budget that covers the work.  Section~\ref{sec:extraction} measures each requirement, and any one of them can fail while the \aone{} edge stays open.

\subsection{Localizing the mechanism with interventions}

We now ask which identity transformation makes the edge appear or disappear on a concrete relay path.  The model of final identities answers through matched interventions that hold the prompt, schedule, and observable fixed while changing one predicted selector:

\begin{enumerate}
\item If an upstream principal or enforced namespace defines the measured lookup domain, splitting the principal or namespace while holding the prompt and schedule fixed should remove the edge between customers while retaining an owner hot repeat.
\item In a credential pool, observations should be determined by the selected partition rather than by the public relay endpoint.
\item In nested relays, an inner overwrite should defeat outer identity preservation, whereas carrying the customer through the inner hop should remove the edge.
\item Within one implementation, adapters that preserve and drop the same field carrying identity should produce different outcomes between customers.
\end{enumerate}

The four predictions localize the transformation responsible for \mzero{} and for observable \aone{} reachability.  How much each result proves depends on its design: a positive endpoint shows exposure on one path and window, while a matched intervention also names the component that controls the edge.

\subsection{From local path evidence to a deployment obligation}

We now ask what identity loss means for a relay serving many mutually distrustful customers.  If the relay maps more downstream security domains into fewer domains enforced by the provider, some customers are necessarily colocated.  This colocation is the structural \mzero{} consequence; the observable capabilities additionally require the conditions set out above.

Counting API keys differs from counting security domains.  Distinct downstream relay keys separate customers only at the first hop, and mapping all of them to one upstream credential provides no separation at the provider.  Distinct upstream key strings become sufficient only where the provider uses those strings to select distinct enforced domains, such as documented separate organizations, workspaces, or projects.  The security quantity is therefore the number of enforced domains the path can obtain.

That quantity is small on the paths we examined.  OpenAI enforces the boundary at the organization and does not sell an additional organization self-serve; Anthropic's direct API draws it one level finer at the workspace, but caps an organization at 100 workspaces and reverts to the organization on Amazon Bedrock and Google Cloud~\cite{openai_prompt_cache,openai_projects,openai_second_org,anthropic_cache,anthropic_workspaces,anthropic_rate_limits}.  Let $D$ be the enforced domains a relay can obtain on a path and $N$ its downstream customers.  Whenever $N>D$, some customers share a final cache identity by counting alone, whatever the operator intends.  A relay above a hundred customers therefore satisfies \mzero{} on these paths through domain supply rather than through misconfiguration.  Appendix~\ref{app:credential-classes} records the documented supply for each credential class, including one class that removes the operator's choice entirely.

The capacity mismatch yields a deployment obligation for every final path.  Every stateful lookup and write must either use a genuinely separate principal enforced by the provider or retain an unforgeable namespace derived from the original downstream customer.  One hop of $\pi$ that substitutes a shared identity reopens the path between customers.


\section{Measurement Methodology}
\label{sec:methodology}

This section defines the measurement process used to identify and explain cache reachability between customers.  We first state the research questions and experimental units, then define the four request unit for a directed cache edge.  We next combine source tracing, runtime observation, and matched interventions to locate the identity transformation, and separately test whether each relay delegates the provider's cache control interface.  We then describe subject selection and frozen execution, introduce the attribution protocol for extraction, and conclude with validity and reproducibility measures.  Table~\ref{tab:measurement-contract} in Appendix~\ref{app:extended-methodology} lists the unit, the controls, and the permitted conclusion for each question.

\subsection{Questions, units, and claim scope}

We ask four questions in order.  \textbf{RQ1} asks whether cache reachability between customers exists on current relay paths.  \textbf{RQ2} asks which identity transformation at the final path or cache control transformation creates, removes, or blocks that edge.  \textbf{RQ3} asks which measured conditions let membership of a known prefix become recovery one token at a time.  \textbf{RQ4} asks which defenses establish separation at the final boundary.

For RQ1 and RQ2, the independent systems unit is a concrete framework or service, route to a model and provider, account pair, dated window, and intervention arm.  The live RQ3 confirmation is one dependent chain with eight positions, while analyses of candidate coverage use the declared conversation, document, fixture, or synthetic path.

\subsection{Classifying a directed cache edge}

Every independent cell instantiates the four request unit of Figure~\ref{fig:a1-protocol}, yielding a strict positive or a controlled miss.

The classifier fails closed.  If the repeat by the owner is cold, the cell is \emph{inconclusive for formation}.  A split in the served model, asymmetric error, retry, or incomparable route is \emph{inconclusive for routing}, and an unreachable inference endpoint is \emph{unassessed for transport}.

One cell therefore establishes a directed edge for a single path and window; the interventions in the next subsection locate its internal cause.

\subsection{Tracing and localizing final identity}

Localization traces the transformation that controls a directed edge.  First, we follow the authenticated customer identity through routing objects, provider credentials, request translation, namespace fields, pools, native and compatibility adapters, retries, and fallbacks to the final provider request.  We retain contrasting paths within a project because one adapter may preserve identity while another drops the field.

For executable gateways, an observer then records digests of authorization and request bodies, fields that carry identity, served model, status, and usage, and verifies the number and order of upstream calls and distinct responses, which rules out a local cache of exact responses.  The observer is what connects source to runtime: source analysis predicts a final identity, the records verify the transformation that actually occurred, and the four request unit measures the cache consequence through a real upstream model with closed weights.

We next hold prompt, schedule, output cap, model label, and observer fixed while changing the identity component the path model predicts.  Shared versus independently scoped provider principals and same versus split namespaces are \emph{controlled interventions}, as are the contrasts between nested overwrite and preservation and between two adapters in one implementation.  Pool outcomes are a \emph{recorded selector association}: we observe which credential partition the gateway selected without assigning it.  Managed versus BYOK moves a whole platform bundle, because credential ownership, capacity, scheduling, and affinity can all change together.  An unmodified production service result is an observation for one route and time.

Each arm tests one counterfactual prediction of Section~\ref{sec:identity-theory}.  When a prediction and its matched outcome agree, we can name the directed edge, the component that controls it, and how strong the resulting causal claim is.

\subsection{Testing cache control availability}

Exposure also needs the relevant cache state to form and stay referenceable, which depends on the interface the relay delegates.  We classify provider interfaces as \emph{automatic}, \emph{opt in by request}, or \emph{explicit object}.  For an explicit object we submit the provider's real creation method through the gateway, send a reference request once creation returns a handle, and record field acceptance, normalized error, upstream availability, usage, and whether authorization depends on the route.  A lifecycle is \emph{rejected at ingress} when a real create request fails before reaching the observer, and \emph{delegated} when creation and the returned reference both traverse the gateway.  Both labels describe the frozen interface, which another interface, authorization grant, or later adapter can change.

\subsection{Subjects, sampling, and frozen execution}
\label{sec:experimental-setup}

We now state how projects, services, and live executions enter the study.  Our open source sampling frame contains 29 retained relay and gateway projects.  Architecture, lineage, adoption proxies, and the availability of a supported mapping from identity to route selected six frozen revisions for source analysis; five were executable.  Before testing a pooled hazard, we verify that each gateway authenticates multiple downstream principals and can constrain one security domain to one exclusive upstream credential, then instantiate both the shared and the exclusive graph.  Provider documentation determines whether distinct upstream credentials select different enforced domains.

Production services use two declared sampling frames.  For OpenRouter, we freeze the union of eligible closed chat model labels appearing in the daily rankings of the top 50 models by token volume from July 26 through August 1, 2026; classification uses only the latest four request protocol and retains untested and inconclusive labels.  For independent relays, we seek at least one eligible instance with two accounts and managed credentials for each executable open source project, while retaining missing strata.  Coded relay names protect ongoing disclosure; the evidence ledger retains family provenance, route to the model and provider, account pair, and time window.

Development, route discovery, and observable calibration are separate from confirmation.  Before a live confirmation, we freeze the account pair, target or canary, prompt digest, schedule, classifier, retry rule, and stopping condition.  We retain the first eligible execution together with every miss and abstention, and the observed security outcome never determines repetition.

\subsection{Protocol for extraction attribution}

We now extend the measurement to dependent extraction.  The live RQ3 protocol starts from an established RQ1 edge and asks when that membership signal can be chained.  The victim writes a known context and then a short secret, so its stored prefix is shorter than the attacker's probe $Pc$ for a candidate extension $c$.  A correct $c$ advances the match one token into the victim's content; a wrong $c$ breaks it at the shared baseline.  Because the probe is longer than the victim prefix, the primary read never reaches the count for a full $Pc$.  The protocol therefore scores each candidate by a relative delta $\delta(c)=R(Pc,V\cup A)-R_{\mathrm{ref}}(c)$, where $R(Pc,V\cup A)$ is the primary read with victim state $V$ and attacker writes $A$, and $R_{\mathrm{ref}}(c)$ is a shadow reference for the same probe.  The \emph{shadow history} is a second context the attacker issues on the same credential, route, and schedule as the primary, with the same structure and the same reported token length, differing only in a fixed length block of pseudorandom marker words at its head.  The victim primes the primary context alone, so the shadow carries the attacker's own writes and no victim state, and reading it removes the prompt length normalization that every candidate shares.  A position commits when exactly one candidate reaches $\delta=\beta+1$ while every other candidate sits at the common baseline $\beta$, an offset of either sign:
\begin{equation}
 \exists!\,c^\star:\ \delta(c^\star)=\beta+1,\qquad
 \delta(c)=\beta\ \ \forall c\neq c^\star .
 \label{eq:relative-selector}
\end{equation}
Attribution to victim state comes from a complete replay that reuses the identical attacker credential, route, and schedule and omits only the victim priming, where every candidate must return to $\beta$.  A delta above $\beta$ present with victim priming and absent in the replay is caused by the victim rather than by the attacker's own writes or by a route that never shared the domain.  Appendix~\ref{app:formal-results} states the provenance argument for comparable and offset floors.

Before any candidate is scored, a control extension outside the candidate set is read on both histories and must return equal counts, which confirms that the primary and the shadow start from the same baseline.  Appendix~\ref{app:extraction-gates} states the numeric thresholds for prompt length, formation, and comparability that every response must then clear; we call those thresholds \emph{gates}.

Before reading a candidate on the primary history, the protocol issues the corresponding shadow request twice.  We call these requests \emph{shadow preheating}: the first creates attacker state for the candidate, and the second supplies the shadow reference used in the candidate delta.  Because the reference is fully formed and the primary stops at the victim's stored prefix, $\beta$ is negative and the commit is the least negative delta.  Shadow preheating covers every candidate because the scorer does not know the truth.  Each primary candidate is then read once in a concurrent batch.  The method commits under Equation~\ref{eq:relative-selector} only when every gate for prompt length, formation, route, status, and retry also agrees, and abstains otherwise.  A fresh pair without victim state then replays the complete dependent schedule and must tie.

The live confirmation freezes one nonce commitment to the target before any traffic and one sequence of eight positions over four words that each form one provider token.  Candidate order is independently shuffled at each position and hidden from the scorer.  The deliberately closed alphabet isolates sequential composition from the quality of candidate generation, which Section~\ref{sec:extraction} measures separately.

\subsection{Validity and reproducibility}

We close with how we protect and reproduce these measurements.  What we measure is whether the reported counts separate one request from another.  Fresh prefixes, owner hot repeats, negative controls, route equality, frozen retry rules, and observer records guard against failed formation, indiscriminate caching, hidden path changes, and local response caching, and any remaining failure is reported as inconclusive.  The sampling frames and dated route windows fix the population and the time each cell represents, and Section~\ref{sec:limitations} collects the broader limits.  Frozen revisions, complete denominators, failure records, and machine readable analyses preserve the evidence behind each reported result.


\section{Evaluation}
\label{sec:causal-evaluation}

This section establishes where current relay paths expose cache state between customers and which transformation controls that exposure.  We first answer RQ1 through configured gateways, OpenRouter, and independently operated production relays.  We then answer RQ2 with interventions on principals, namespaces, pools, adapters, and nested relays.  The last subsection tests whether each relay delegates the cache control interface that state formation requires, and closes with the answers to RQ1 and RQ2.  A cell \emph{crossed} when one customer read state written by the other.  Table~\ref{tab:main-results} maps this section, one row per finding; Appendix~\ref{app:extended-methodology} contains the complete evidence ledger and classifiers.

\begin{table*}[t]
\centering
\caption{Core evidence supporting the answers to RQ1 and RQ2.}
\label{tab:main-results}
\vspace{3pt}
\small
\setlength{\tabcolsep}{4pt}
\renewcommand{\arraystretch}{1.18}
\begin{tabularx}{\textwidth}{@{}p{2.75cm}Y@{}}
\toprule
Finding & Evidence and security implication \\
\midrule
Gateway capability & All five executable projects authenticate multiple downstream principals and can be configured to reach an exclusive upstream credential, but ordinary selection uses a group, provider, route, deployment, or account pool, so avoiding pooling takes an explicit isolation policy. \\
Two upstream families & Deliberately shared configurations crossed in all five OpenAI cells ($[0,1408,1408,0]$) and all five native Anthropic cells ($[0,h,h,0]$, with $h$ from 1,066 to 1,074), so both cache lifecycles exposed membership of a known prefix across distinct relay keys. \\
Path localization & Principal and namespace splits removed their matched edges; recorded choices of the same pool partition crossed 3/3 and different partitions 0/3; an inner overwrite defeated outer separation.  Reachability is determined by the identity the final path selects. \\
OpenRouter ranked frame & Among 28 eligible labels in a frozen weekly ranking by token volume, strict evidence covered 80.5\% of listed eligible volume: 12 labels had a positive route, one a negative on its tested path, 11 were inconclusive, and four remained untested.  Reads across accounts recur among labels with high token volume. \\
Relays stratified by framework & Eligible deployments with two accounts were found for two of the five executable projects, yielding four deployments.  Three of the four had a positive edge; the fourth had one controlled negative direction and one inconclusive direction.  Independent operators confirm recurrence in the strata available. \\
Cache control interface & OpenAI automatic state and Anthropic state enabled by a request were available through all five pooled gateways, while four gateways rejected a real Gemini cache creation request and LiteLLM forwarded creation and reference after an explicit authorization grant.  Exposure therefore depends on both final identity and delegation of the provider's cache lifecycle. \\
\bottomrule
\end{tabularx}
\end{table*}

\subsection{RQ1: Does exposure exist on current relay paths?}

\paragraph{Gateways configured by researchers.}
We audited NewAPI~\cite{newapi_repo}, One API~\cite{oneapi_repo}, uni-api~\cite{uniapi_repo}, MetaAPI~\cite{metapi_repo}, LiteLLM~\cite{litellm_repo}, and Sub2API~\cite{sub2api_repo} at frozen revisions.  One API supplies source and lineage evidence.  Each of the other five authenticates multiple downstream principals and can express a graph with an exclusive upstream credential, through a group and channel, a rule qualified by provider, a policy restricted to one route, a deployment scoped to one team, or a group holding one account; Appendix~\ref{app:extended-methodology} maps each object to its project, and no project documents an enforced domain per downstream key.

\paragraph{Shipped defaults.}
Expressing that exclusive graph is an operator action, so we also determined what each project does before an operator takes it.  At the frozen revisions none of the five selects an upstream credential using downstream customer identity: selection is keyed by model, group, provider, or route, and the shipped defaults spread requests across every credential that can serve the model.  Three of the five reselect a credential on every request, so one customer's prompts reach every credential's cache domain over time, and adding credentials increases rather than reduces the domains holding that customer's content.  NewAPI's retry additionally descends priority tiers, so a separation an operator builds from priority is crossed on retry.  Appendix~\ref{app:default-posture} records 21 falsifiable source probes and the mechanism for each project.

We executed all five with two downstream principals mapped to one upstream credential controlled by the researchers, which reproduces the shipped default under the domain supply of Section~\ref{sec:identity-theory}; Appendix~\ref{app:default-posture} establishes that default.  On OpenAI, the observer saw exactly four upstream requests in $P/P/P/P'$ order, one authorization digest, the expected request digests, and distinct responses; every gateway produced $[0,1408,1408,0]$.  We repeated the matrix through each gateway's native Anthropic Messages ingress with an explicit cache breakpoint.  All five again produced $[0,h,h,0]$, where $h$ ranged from 1,066 to 1,074, and provider usage separately recorded cache creation for both the prime and the isomorphic fresh negative.  The resulting ten cells connect the source prediction to runtime across five gateways and two cache lifecycles.

\paragraph{OpenRouter.}
With account pair, model label, provider ordering, prompt, and protocol fixed, OpenRouter managed capacity crossed 5/5.  Independently scoped BYOK crossed 0/5, retained 5/5 owner hot repeats, and kept all five fresh negative responses cold.  The 5/5 managed and 0/5 BYOK contrast reproduces CacheProbe's closest production result~\cite{fahey2026cacheprobe}.  Managed and BYOK jointly change several platform components, so the intervention identifies the production contrast at the level of that bundle.

A competing reading of any single positive is that the upstream provider serves one cache to all of its own customers, as Gu et al.\ report for several production APIs, in which case the observation would say nothing about what the relay does with customer identity~\cite{gu2025auditing}.  An independently scoped credential separates the two readings, because a globally shared upstream cache still crosses when each customer holds a distinct credential.  BYOK crossed 0/5 on this route, so the shared upstream credential is what produced the managed positive.  The direct service matrices below measure the same boundary at two providers with no relay in the path, and reach the same conclusion for those families.

The weekly frame of Section~\ref{sec:experimental-setup} contains 28 labels ordered by listed volume, every one of them selected before any cache outcome was known.  Cells from the latest protocol cover 80.5\% of the frame's listed token volume for eligible models: 12 labels have at least one strict positive route, one has a negative on the tested path, 11 are inconclusive, and four remain untested.  Labels with a positive route account for 7.1\% of OpenRouter's platform denominator and 33.7\% of the volume for eligible models.  The ledger retains duplicates and discordant cells.  We ran the credential class contrast on one route rather than on each label, so this frame shows that reads across accounts recur among labels carrying high token volume, and the next subsection asks where identity is lost.

\paragraph{Additional production relays.}
Coded services supply deployment evidence from upstream mappings operated independently of our study, and availability decided which ones we could include: four attributed deployments in two of five strata, with the uni-api, MetaAPI, and LiteLLM strata missing.  Two NewAPI deployments gave positive GPT and Grok paths and one controlled negative direction with an inconclusive reverse; two Sub2API deployments gave sparse strict positives, 2 of 9 and 1 of 8 valid cells on their default routes.  Two further services of unidentified lineage each produced bidirectional $[0,1024,1024,0]$ positives.  The same behavior therefore recurs across independent operators and lineages.

\subsection{RQ2: Where is identity lost?}

\paragraph{Principal and namespace.}
The RQ2 interventions now locate the identity transformation that controls exposure.  Some of these interventions run at a \emph{direct service}, by which we mean a model provider reached with our own credentials rather than through a relay.  A direct service measures what the provider itself uses to select a cache domain, which is the premise every relay result rests on, because a relay can only erase a distinction the provider enforces.  We denote the three direct services we tested Provider~A, Provider~B, and Provider~C.

At Provider~A, a matrix varied the upstream principal and the request namespace independently over 15 trials assigned at random, and all 15 scheduled trials were valid.  One principal with one namespace crossed 5/5, splitting the namespace under one principal crossed 0/5, splitting principals while reusing the same namespace value also crossed 0/5, and every later repeat by the owner remained hot.  Either variable alone therefore selects the measured lookup domain, so a relay that shares either value across customers merges those customers into one domain.  Each split arm against the shared arm is a randomized comparison of five trials against five, whose exact two sided value is $p=0.008$, the smallest value that size can produce.  The namespace field carries no documented access control guarantee, so it delivered a measured lookup domain while the enforced domain remained the provider principal.  For NewAPI and LiteLLM, changing from one shared provider principal to two independently scoped principals changed two strict positive cells to two controlled misses: $[0,1408,1408,0]$ became $[0,0,1408,0]$.

Within one Sub2API build, the generic OpenAI adapter forwarded the shared caller cache key and crossed at $[0,185,185,0]$, while the native Grok adapter derived a per customer namespace from the downstream key and produced a controlled miss at $[0,0,184,0]$ with the owner repeat still hot.  The same intended identity therefore survived one adapter and disappeared on another inside a single project, which shows why a project name cannot serve as a security label.  A comparison across projects agreed: a NewAPI path that preserved the namespace remained separated where LiteLLM's Chat adapter dropped the field and crossed.  Provider~B produced a controlled miss across workspaces while the repeat in the priming workspace remained hot.  At Provider~C, two teams verified through the API shared a fresh identifier for conversation and routing and showed an increment of 832 cached tokens over a background of 128; we record this as an association, because the experiment did not vary that identifier.  Providers~A and~B therefore separate domains by a value a relay controls, whereas Provider~C already shares one across teams before any relay is involved.

\paragraph{Pools and nested relays.}
Pools and nested relays test whether later routing can replace the direct selectors above.  In a randomized pool with two keys, every choice of the same partition crossed (3/3, 95\% interval $[0.292,1]$) and every choice of different partitions produced a controlled miss (0/3, $[0,0.708]$).  Observations were determined by the recorded final credential partition rather than by the public relay identity, so pooling changes the collision schedule and an adaptive retry can revisit a shared partition.

The nested experiment held prompt, accounts, credential, model, endpoint, observer, and schedule fixed across three arms and contributed one cell per arm against a real upstream.  A shared path crossed, an inner hop that overwrote the outer customer identity also crossed, and a hop that preserved the original customer removed the observation between customers while retaining an owner hot repeat.  One question those cells leave open is whether each verdict is determined by the identity the chain forwards or by the particular prompt used.  We rebuilt both gateways from their frozen revisions and replayed the arms over five fresh canaries against a local fixture that records the namespace it receives.  Every arm forwarded the namespace its source predicted on all five canaries: one value shared by both customers on the shared and inner overwrite arms, distinct values on the preserved arm, and no verdict changed with the prompt.  Because that fixture keys its cache on the namespace it receives, the replay verifies what the gateways forward, while the real upstream cells carry the provider evidence.  The nested and adapter contrasts together confirm the compositional result: a later transformation that redefines the final identity defeats separation at an outer hop.  Appendix~\ref{app:extended-methodology} reports the exact interval for every arm.

\subsection{When does cache authority extend to the relay?}

A shared identity becomes observable state only if the relay delegates enough cache authority.  Table~\ref{tab:provider-compatibility} separates the provider cache lifecycle from the interface delegated by each evaluated path.  On compatible OpenAI paths, ordinary eligible inference formed automatic state and produced reads between customers.  Anthropic required explicit cache control in the request: enabling cache control incurred the documented creation charge and allowed a later read between customers, which confines the affected workload to requests that enable caching~\cite{openai_prompt_cache,anthropic_cache}.  Gemini's implicit cache managed by the provider supported ordinary inference, and every valid tested cell reported no read between customers.

Gemini uses a separate control interface for its explicit cache, so we submitted a real \texttt{POST /v1beta/cachedContents} creation request instead of inferring support from ordinary generation.  NewAPI, uni-api, MetaAPI, and Sub2API rejected it at the tested ingress with 404 or 405, before the request reached the observer.  LiteLLM's generic Google interface gave the matched contrast: a virtual key without route permission received 403, while granting \texttt{/gemini} forwarded creation and the returned resource reference to the local fixture.  Availability of the explicit cache therefore depends on the adapter and its route permission, and because explicit storage bills over time, that permission also decides who pays~\cite{google_gemini_cache,google_gemini_pricing}.

Both questions now have answers.  Strict positives on controlled and production paths establish a membership channel for a known prefix between customers, and the provider lifecycle decides whether a shared identity creates state another customer can reach.  Principal and namespace splits, recorded pool partitions, adapter contrasts, and nested overwrites then locate the transformation in control, so every evaluated path either preserves identity, collapses it into reachable state, or maps the request to a cache feature the relay never delegates.


\section{Evaluating the Extraction Capability Limit}
\label{sec:extraction}

\begin{figure*}[t]
  \centering
  \includegraphics[width=\textwidth]{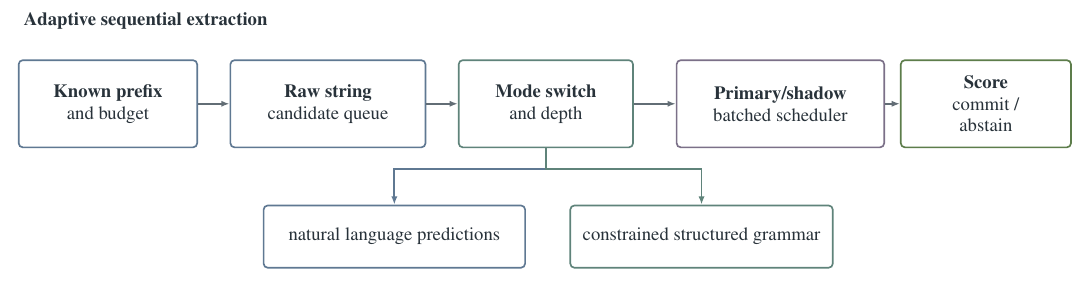}
\caption{The extraction pipeline evaluated in this section.  A known prefix and a request budget feed a queue of candidate strings, and a mode switch draws them from natural language predictions or from a constrained grammar.  The scheduler reads each candidate once on the primary history and twice on the shadow history of Section~\ref{sec:methodology}, and the frozen score commits one candidate or abstains.  Only that score decides whether a lift belongs to the victim.}
  \label{fig:extraction-pipeline}
\end{figure*}

This section measures when a cache report can reveal content beyond a known prefix.  We first test whether attributable feedback recovers an unknown sequence through an unmodified production relay.  We then measure how candidate breadth, an unknown tokenizer, reporting granularity, route integrity, and protocol cost control broader extraction.  The final subsection combines the production and local results into the capability limit supported by RQ3.  Figure~\ref{fig:extraction-pipeline} shows the pipeline these measurements exercise.

\subsection{Recovering a controlled unknown sequence}

The controlled trial from Section~\ref{sec:experimental-setup} committed one target sequence of eight positions before traffic, and a controlled victim account created the target cache state.  At each position, the scorer tested four known words that each form one provider token on the served \texttt{grok-4-fast-non-reasoning} path.  The scorer had no access to the target word.  All eight dependent commits were correct with no abstention.  Each committed word became the only prefix used at the next position, so one wrong decision would have corrupted the remainder.  The target space contained $4^8=65{,}536$ sequences, which represents 16 bits of committed target rather than a bound on the method.

The victim wrote a known context of about 960 words and then the secret, so the victim prefix was shorter than each attacker probe.  Every probe matched that prefix up to the point of divergence: a wrong candidate stopped at the shared baseline, and the correct candidate advanced the match one token further.  Scored against the shadow reference for the same probe, the three wrong branches at the first position sat at $\beta=-15$ while the true branch reached the unique $\beta+1=-14$, and the same adjacent lift recurred at all eight positions; Table~\ref{tab:stagec-raw-position} lists every read at that position.  The complete replay reused the identical attacker credential, route, and model and differed only in that the victim did not prime; there every branch tied at $-15$, so victim priming caused the lift, not the attacker's own writes and not a route that never shared the domain.  All gates for model, status, prompt length, formation, route, and the reuse threshold passed.  The decisions that included victim state used 128 requests on the attack path; two victim formation requests and the 128 request replay without a victim bring the total to 258, which Appendix~\ref{app:extraction-gates} decomposes.  Appendix~\ref{app:extraction-scope} provides the full request, token, and timing ledger.

The scorer made 32 logical candidate decisions; attribution and integrity raised the controlled experiment to 258 HTTP requests.  We retained the target commitment and its opening, the candidate order, all sanitized usage rows, the matched replay, and an offline verification.  The commitment manifest is the first record in the frozen request log, establishing that the target was fixed before any traffic.

\subsection{Which conditions control broader extraction?}

The controlled chain establishes recovery of 16 bits under favorable conditions; broader extraction depends on how often those conditions hold.  A candidate response is \emph{cold on its route} when the known prefix for that candidate returns $R(q)<\tau(q)$, which leaves that branch without the common cached prefix needed to compare its score against the others; expiry and a different route selection look the same from outside.  We denote one further coded production service Relay N, and for candidate coverage we use conversations from the public WildChat dataset and the public dictionary of 2,048 mnemonic words defined by Bitcoin Improvement Proposal 39 (BIP-39).  Table~\ref{tab:a2-factors} reports the measured factors together with the route condition.

\begin{table*}[t]
\centering
\caption{Measured factors governing whether \aone{} feedback can become \atwo{} extraction one token at a time.}
\label{tab:a2-factors}
\footnotesize
\setlength{\tabcolsep}{3.5pt}
\renewcommand{\arraystretch}{1.10}
\begin{tabularx}{\textwidth}{@{}p{2.35cm}Y@{}}
\toprule
Factor & Measurement and capability implication \\
\midrule
Cache formation and control & OpenAI automatic state and Anthropic state enabled by the request were available through relays; four gateways rejected a real Gemini create request (Table~\ref{tab:provider-compatibility}).  A rejected ingress blocks only the tested path: another authorization or adapter can delegate the lifecycle. \\
State reachability between customers & Strict production \aone{} positives coexist with a separate Relay N route that has 2/2 bidirectional hits and no attributable oracle for the first extension.  A hot response for the known prefix supplies the starting signal; extraction additionally needs discrimination among extensions. \\
Reporting resolution & The live path exposed eight adjacent $+1$ lifts.  On 192 BIP-39 positions with the correct previous words supplied, identifiable counts fell from 192 at one token granularity ($g=1$) to ranges of 96 to 97, 48 to 49, and 16 to 17 at $g=2,4,8$, so coarser reporting sharply reduces identifiable positions.  Across eleven Grok labels on one relay, one reported strict one token resolution and two reported a coarse 128 token block, so resolution belongs to the route (Appendix~\ref{app:extended-methodology}). \\
Attribution despite cache writes & Primary and shadow histories plus replay without a victim separated victim lift from attacker writes, whereas repeating an asymmetric side could erase provenance.  Attribution sets the error a defensible extraction claim accepts. \\
Route and residence & Two live runs with eight candidates retained the unique truth lift at all seven evaluated positions, but one wrong candidate response was cold in each run and forced abstention.  Route erasures block composition even where the extension signal survives. \\
Candidate coverage & Across 76 WildChat source conversations under three tokenizer conditions, both arms reconstructed the span ending at the same absolute token, and one known starting token raised the fraction of records that recovered it from 18.9\% to 64.5\% at endpoint 9 and from 11.4\% to 51.3\% at endpoint 17.  Without target token identifiers, a public dictionary of 2,048 words completed all 16 local paths with twelve positions at $g=1$ under each of three unknown tokenizer conditions.  Public frontiers and structured dictionaries can turn generation into enumeration. \\
Budget & The live proof used 258 controlled requests.  The same integrity protocol requires 98,370 requests for eight positions with 2,048 candidates, or 136,034 for a valid BIP-39 schedule with twelve words, so rate, token, residence, and anomaly budgets control current feasibility. \\
\bottomrule
\end{tabularx}
\end{table*}

\paragraph{Attribution, route, and coverage.}
Exact local symmetry between primary and shadow produced no wrong commits.  We then screened 46 local paths, each requiring a median of 37,876 to 45,710 paired shadow calls.  Under an assumed error model rather than a production measurement, with an independent differential error of $10^{-5}$ per comparison, the mean chance of selecting a wrong candidate at some position on those paths is 22.59\%, which is why the protocol abstains whenever a gate fails.  Route integrity failed independently of candidate breadth: a separate Relay N route produced six jointly hot frontiers yet no paired symmetry for the first extension in 2 cells, because attacker repetition created reusable state on both sides.  Coverage with an unknown vocabulary reached from 88.7 to 89.8\% across 16 documents drawn from a subset of the public \texttt{openai/coval} comparison corpus pinned by hash.  Appendix~\ref{app:local-studies} reports the remaining local measurements.

\subsection{Answer to RQ3: measured extraction limit}

Those measured factors explain both the successful chain carrying 16 bits and the wider runs that abstained.  The local result over a full dictionary shows that an unknown provider vocabulary still permits structured enumeration, while the live runs with eight candidates show that route integrity can fail before candidate space becomes the dominant constraint.  RQ3 therefore has a measured answer: one production route supported attributable recovery of a controlled unknown sequence, and the remaining experiments identify the conditions that govern broader recovery.


\section{Deployable Defenses}
\label{sec:defenses}

The measured path failures imply defenses that a relay can deploy.  We first rank available mechanisms by the guarantee they provide and name the primitive a provider would have to expose for a relay to attain the strongest level.  We then specify the defense contract for every final cache boundary and validate routing and failure behavior across the five executable gateways.  Next, we place the split after reusable public prompt material to recover most caching benefit, and the final subsection evaluates reporting granularity and complementary controls that raise attack cost.

\subsection{Guarantee hierarchy and observable surface}

Strict isolation, as Section~\ref{sec:background-threat} defines it, requires every transcript the attacker can see, including usage, timing, billing, balances, headers, errors, quotas, routes, and future state, to be independent of another customer's private cache history.  Four mechanisms are available, and they differ in what they guarantee.  A separate principal, workspace, subscription, or project enforced by the provider gives the clearest deployed separation, and it is the preferred defense wherever enough enforced domains exist.  An opaque namespace enforced by the provider separates cache state whenever the namespace reaches every lookup and write; vLLM's request field \texttt{cache\_salt}, which carries security semantics, shows that a serving stack can implement it~\cite{vllm_cache_salt}. However, no provider in our measurements offers the namespace primitive to an API customer.  On those paths the strongest defense a relay can build alone is a separate provider principal, and Section~\ref{sec:identity-theory} bounds how many principals it can obtain.  Closing the remaining gap needs providers to expose an enforced namespace; the next subsection states what a relay can do with what exists today.  Table~\ref{tab:defense-guarantees} in Appendix~\ref{app:extended-methodology} records the claim each primitive supports.

\subsection{Defense contract at every final boundary}

We now specify how a trusted relay derives an enforced identity from the authenticated original customer.  We use a keyed hash message authentication code (HMAC) in the following construction:
\begin{equation}
 n_d=\operatorname{Trunc}_{k}\!\left(
 \operatorname{HMAC}_{K}(\texttt{tenant/v1}\parallel
 \operatorname{LP}(d,m,v,\varphi))\right).
 \label{eq:namespace}
\end{equation}
$K$ is unavailable to customers; encoding each field with its length prevents ambiguous concatenation; $v$ identifies the provider or deployment and $\varphi$ the policy version.  The relay ignores values supplied by clients for access control and fails closed on canonicalization errors.  Rotation uses explicit versions and permits lookups only under a declared namespace.

Our reference configuration binds the model $m$ and the provider or deployment $v$ into the namespace as well, which keeps a customer's entries from being reused across deployments that tokenize or cache differently.  Truncating to $k=128$ bits leaves a negligible chance that two customers receive the same namespace; Appendix~\ref{app:formal-results} gives the union bound an operator recomputes for its own population and rotation window.

The contract is universal over the path: $n_d$ must survive provider selection, pools, aliases, streaming, retry, fallback, and nesting, and the final runtime must enforce it for both lookup and write.  Given enforcement and no active collision, entries from $d\neq d'$ then occupy disjoint lookup and write domains while the same $d$ keeps its reuse.  That is conditional nonreachability of cache state; isolating the full transcript additionally requires equality of billing, timing, compute, errors, routes, quotas, and future state.

\subsection{Gateway and upstream validation}

We now test whether each gateway can realize the required path behavior during success and failure.  Table~\ref{tab:defense-exact-vectors} executes shared and exclusive credential graphs against a local fixture that treats canonical credential identities as cache domains.  Shared paths crossed in 5/5 gateways.  Exclusive paths crossed in 0/5 and retained all five owner hot repeats.  The exclusive results verify that each configured graph routes A and B into disjoint credential identities, which is what this fixture measures.  Whether those identities are also distinct enforced domains is a separate premise set by the provider: on OpenAI two credentials in one organization share the cache, so exclusive credential separation requires distinct organizations, which Section~\ref{sec:identity-theory} shows are not self-serve.

Namespaces selected by clients provide weaker assurance.  Honest distinct values blocked sharing in 4/5 paths, while LiteLLM Chat dropped the field, and choosing the same value restored sharing on every shared path where the field could matter.  Our reference arm, NewAPI's HMAC namespace derived by the relay, overwrote the forgery, removed the edge, and retained reuse by the owner.

We next injected a deterministic 503 for B's exact credential presented to the provider without admitting cache state.  Across the five exclusive graphs, A's requests remained usable and its repeat remained hot, while B returned an error or was stopped locally, and every later B attempt stayed outside A's credential.

With a real OpenAI upstream, NewAPI shared and forged client arms each hit across customers in 3/3 trials with $[0,1024,1024,0]$.  The relay HMAC arm produced 0/3 hits across customers and 3/3 hits for the owner with $[0,0,1024,0]$; all 36 requests reached the same served model.  Because only the routing key changed, that intervention identifies the cause of the measured lookup change.  The provider documents the field for bucketing rather than for access control, so today the result is a conditional lookup mitigation, and a provider security guarantee would lift it to the namespace tier.

The two results above use a real upstream.  The remaining checks are \evidence{E1}: they run against a model of the gateway rather than a provider.  We tested properties over 50,000 sampled local gateway transcripts containing failures before and after writes, partial streams, retries, quotas, refunds, costs, routes, and future state.  The policy selected no weaker fallback and disclosed no marker; 8,389 cases, about a sixth, failed closed, and the sampled model produced no counterfactual difference.  A separate simulator found equality in 18/18 cells with stable customer identity and kept private reuse for the same customer in 18/18 cells.  Those 18 cells enumerate the observables the simulator models, so they show the policy is internally consistent over that surface and leave anything outside it untested.

\subsection{Preserving reuse of public prefixes}

We next cut the contract's cost by keeping reuse for the long public preambles that agents, systems, and tools send.  A \emph{public/private split} shares only a versioned public region the operator has verified, and introduces $n_d$ before the first private token.  The operator registers immutable public templates; system instructions specific to a user, retrieval, tool output, memory, environment metadata, and dynamically inserted template content are private by default.  Each decision is determined by template provenance, and labeling dynamic content as public violates the configuration rule.

We evaluated the design by replaying 5,713 request events from 3,000 sampled WildChat conversations after five public agent prefixes ranging from 1,553 to 8,523 \texttt{o200k\_base} tokens, under concentrated, weighted, and balanced prefix mixtures.  The replay persisted no raw conversation or identifier.  The cache floors of 256 and 1,024 tokens, the 128 token reporting block, and the price multipliers of 0.1 for reads and 1.25 for writes are representative modeling parameters tested for robustness rather than one provider's published tariff.  Under them the split increased modeled cost by 1.7 to 2.5\% and exposed zero cached tokens from private regions across customers.  Placing the split is not enough to reach that zero.  The replay reports reuse in blocks of 128 tokens, so a split landing inside a block would publish the private tokens sharing it.  The placement rule therefore has to align with the reporting block at every admission floor.  Complete namespacing instead cost from 2.29 to 3.14 times the unsafe baseline, because full separation gives up the shared agent preamble reuse that the split keeps.  Anthropic further excludes cache reads from the input token rate limit, so full separation also forgoes throughput headroom on that provider~\cite{anthropic_rate_limits}.  These figures count tokens only and omit the availability cost of failing closed.

A visible HMAC prompt marker is the weaker fallback for a relay with no other place to put the namespace.  A marker of 96 to 128 bits consumed a median of 17 to 24 tokens across tested tokenizers, and in fixed local task families a baseline of 100/100 fell to 79/100 and 85/100 for two marker variants.  A pilot of 48 requests to a model with closed weights saw no regression the marker alone caused, so the risk is compatibility rather than correctness, and it still favors opaque provider metadata or an explicit split after public content.

\subsection{Granularity and defense in depth}

We now consider controls that instead reduce the utility of the observable.  Downward quantization $Q_g(z)=g\lfloor z/g\rfloor$ hides most gains of a single token, yet a correct next token still becomes visible whenever the extra matching token crosses the next reporting boundary.  Alignment or a known matching suffix can produce that crossing even at $g>1$, so isolation keeps depending on identity.  Appendix~\ref{app:formal-results} bounds the success probability of an attacker testing $B$ block candidates when every unknown provider token carries pointwise conditional minimum entropy $\eta$.  Templates, code, checksums, public suffixes, and explicit grammars supply a defensible $h$; natural language does not, because average surprisal gives no pointwise floor, which leaves granularity as defense in depth there.

Secondary controls include saturating reporting after the split, delayed cohort accounting independent of cache state, calibrated timing padding, query budgets across accounts, shorter residence, classifiers for sensitive regions, and anomaly detection among sibling prefixes.  SafeKV, CachePrune, and PrefixWall preserve sharing through classification or policy aware of ownership, and they complement identity at the final boundary with assurance set by their detector and policy coverage~\cite{chu2025safekv,wu2026cacheprune,pennas2026prefixwall}.

RQ4 therefore has two layers.  Cache state becomes unreachable across customers once every final lookup and write carries either an unforgeable namespace for the original customer, with a bounded collision probability, or an independently isolated provider principal; the controls above only change how expensive one extraction attempt is.  Isolating the full transcript additionally requires equality of billing, timing, compute, errors, routes, quotas, and future state.  An operator attains the first layer by inventorying every endpoint, transport, pool, alias, retry, fallback, and nested path, deriving the namespace from authenticated identity, testing access across customers, reuse by the same customer, forgery, and failure, and failing closed wherever propagation cannot be established.

\section{Discussion}
\label{sec:discussion}

Separate authentication and billing lose their security meaning when a later relay path presents two customers as one cache identity, so an upstream credential can make a relay a security principal even when the relay never stores a prompt.  Because the supply of domains enforced by providers is bounded, a large relay chooses which customers to colocate rather than whether to pool, and inexpensive virtual namespaces enforced by providers would remove the pressure.  Gateways that redistribute consumer subscription quota hold credentials exposing no workspace, project, or administrative interface, so their pooling follows from the product rather than from a setting an operator can correct.

The membership channel is anonymous within a cache domain, yet its resolution sharpens toward a specific customer when the shared domain is small or the customer set is otherwise bounded, so an operator should treat a narrow pool as a stronger disclosure than a large one.  The same primitive admits a defensive use: an operator that plants a unique canary in its own protected prefix can detect the prefix reaching a domain it should never reach.

An aggregate hit rate for a service cannot identify whether exposure comes from a principal, namespace, pool, adapter, or nested overwrite, and an aggregate miss cannot distinguish isolation from feature rejection or route mismatch.  Provider telemetry for the final principal, namespace, pool partition, cache formation, and retry or fallback route would allow exposure estimates beyond the reach of customers who see only a black box.

\section{Limitations}
\label{sec:limitations}

\subsection{Sampling and prevalence}

The OpenRouter label frame is selected independently of outcomes but does not estimate prevalence at the request level.  Its ranking omits credential class, route shares for each provider, eligibility, and cache outcomes for individual requests.  The frame also uses one account pair, so a label classified positive reflects a single draw of the pool partition that determines the observation in Section~\ref{sec:causal-evaluation}; separating a label property from a particular draw needs independent account pairs.  The production relay frame is limited by availability, with three of five framework strata missing.  Both frames strengthen recurrence and guard against selective reporting, while latent routes, pool weights, and residence distributions preclude rates over instances, users, or affected traffic.  Appendix~\ref{app:ecosystem-denominators} preserves the incompatible ecosystem denominators.

\subsection{Extraction external validity}

We found no stable production setting for general extraction of arbitrary content.  The strongest evidence is one chain with eight steps over four known words of equal token length: 128 requests on the victim path and 258 total requests demonstrate composition in one favorable cell carrying 16 bits.  This experiment does not measure route reliability, practical speedup, or recovery of natural language, credentials, or content from a third party.  Its one token reporting was calibrated on a different Grok label than the served chain, and because granularity varies by route across that family, the chain depends on its own gates rather than on a quantum shared by the family. Local paths over the complete dictionary establish enumeration without provider token identifiers only under an exact oracle, and their extrapolations from 98,370 to 136,034 requests establish no production residence, reliability, cost, or detectability.  Full accounting and historical eligibility appear in Appendix~\ref{app:extraction-scope}.

\subsection{Defense and observable coverage}

The routing and failure matrix covers five frozen local gateway graphs, while the namespace derived by the relay has one gateway and one lookup validation with a real upstream.  The local fixture models credentials as cache domains; real deployment additionally requires a principal, workspace, or project boundary documented by the provider.  Trace costs omit provider capacity effects, and finite sampling of failure transcripts supplies test coverage rather than proof.  Namespace separation remains conditional on unforgeable derivation, collision avoidance, and propagation to every lookup and write; full transcript isolation additionally covers billing, timing, compute, errors, routes, quotas, and future state.  Timing lies outside our core evidence: the retrospective set without streaming is near chance and lacks randomized controls for time to first token.

\section{Related Work}
\label{sec:related-work}
Prior work establishes three foundations for our study: production cache sharing, reconstruction from cache observables, and controls on cache reuse.  Our study connects those foundations at the identity transformation performed by an API relay.  We organize the comparison by the question each published design answers: endpoint exposure, content reconstruction in a serving environment, or localization of the relay component that creates the final cache boundary.

\subsection{Gateway cache isolation}

CacheProbe compares paths through a direct provider, managed OpenRouter, and independently scoped bring your own key (BYOK), observes reuse across accounts on managed capacity, and attributes the contrast to gateway credential sharing, which establishes production exposure at the endpoint level~\cite{fahey2026cacheprobe}.  Managed versus BYOK changes principal ownership, capacity pool, scheduler, route, and affinity as one bundle, so that contrast leaves the controlling transformation unidentified.  Our study separates the components and predicts outcomes for paths containing multiple credentials, adapters, pools, and nested relays. \systemname{} traces candidate transformations in various sources, verifies the identities leaving five gateways through observers attached to real upstream models, and tests counterfactuals for principals, namespaces, recorded pools, adapters, and nested hops.  Requests crossed only on the same selected pool partition, and an inner overwrite defeated an outer namespace that otherwise separates customers.  Provider compatibility adds a second dimension to credential audits, because a shared identity exposes automatic state, exposes state enabled by a request only for marked requests, and reaches an explicit cache only through a delegated control interface.

\subsection{Provider audits and cache side channels}

Gu et al.'s \emph{Auditing Prompt Caching in Language Model APIs} uses statistical timing tests to distinguish caching and sharing levels across production providers~\cite{gu2025auditing}.  Early Bird studies timing channels in shared key and value (KV) caches and semantic caches, and InputSnatch combines timing with learned candidate generation~\cite{song2025earlybird,zheng2024inputsnatch}.  The studies establish that observables dependent on cache state can reveal state across users.  Gu et al. characterize sharing at the provider boundary, whereas our measurement follows an authenticated relay customer through the identity transformations that precede that boundary.  Our primary observable is the integer count of cached input reported by the provider, which keeps network timing out of the core experiment and leaves billing, routing, errors, and future state as further surfaces.

The broader systems pattern predates LLM serving.  Confusion over web cache keys shows how an intermediary can omit a request component relevant to security, and deduplication across virtual machines shows how a shared optimization can create a signal between principals~\cite{mirheidari2020cachedconfused,bosman2016dedup}.  LLM relays add authenticated schema translation, pooled provider credentials, nested routing, cache authority specific to each provider, adaptive cache writes, and long public prefixes before private content.  The relay features make the security boundary a composition of identities and interfaces that control state.

\subsection{Prompt reconstruction}

PROMPTPEEK reconstructs cached prompts through scheduling based on the longest prefix in white-box systems; Early Bird also reconstructs cached prefixes through timing; InputSnatch and OptiLeak improve candidate construction; and SpliceLeak studies KV fusion beyond prefixes in retrieval augmented generation (RAG)~\cite{wu2025promptpeek,song2025earlybird,zheng2024inputsnatch,wang2026optileak,sun2026spliceleak}.  PROMPTPEEK provides reconstruction in a locally controlled serving runtime, while our extraction contribution addresses attribution and dependent recovery through a production relay with closed model weights. In our settings, the raw string interface operates without provider token identifiers.  Primary and shadow histories distinguish victim state from cache writes by the attacker, and a replay without victim state validates dependent attribution.  The production and local results together define a measured capability limit for recovery through a real relay.

\subsection{Cache sharing defenses}

SafeKV, CachePrune, and PrefixWall preserve selected sharing through block safety, filtering of sensitive tokens, or ownership policy~\cite{chu2025safekv,wu2026cacheprune,pennas2026prefixwall}.  Assurance for the three schemes depends on classification, policy coverage, and the first probe.  Our requirement complements these controls: an unforgeable identity for the original customer must be present at both the final lookup and the final write on every path.  The serving primitive already exists in vLLM and SGLang~\cite{vllm_cache_salt,sglang_radix}, so what remains is for a relay to propagate and enforce the cache identity across native and compatibility endpoints, pools, retries, fallbacks, and nested hops.


\section{Conclusion}
\label{sec:conclusion}

Prompt cache confidentiality at a relay belongs to the concrete final path: separate relay keys protect customers only while every later hop preserves the identity that keys the provider's cache.  \systemname{} locates where that identity is lost, by combining source tracing of candidate transformations, runtime records of the request that reaches the provider, and matched interventions that name the controlling principal, namespace, pool, adapter, or nested hop. Measurements across five executable gateways, two provider cache lifecycles, OpenRouter, and independently operated relays establish recurring reachability between customers, and one production route additionally supported attributable recovery of a controlled target carrying 16 bits.  The remedy follows from the same analysis.  Every customer must enter a domain the provider enforces, or a namespace derived from authenticated customer identity must survive every final lookup and write, including retry, fallback, pool, adapter, and nested relay.  A split after reusable public prompt regions then preserves most shared reuse.

\bibliographystyle{IEEEtran}
\bibliography{related_work,project_sources}

@misc{newapi_repo,
  author = {{QuantumNous Contributors}},
  title = {{New API}},
  howpublished = {GitHub repository},
  url = {https://github.com/QuantumNous/new-api},
  year = {2026},
  note = {Evaluated commit e0d5156115881780328d31fe9bce7fe25aa9c6c7 (2026-07-20)},
}

@misc{uniapi_repo,
  author = {{uni-api Contributors}},
  title = {{uni-api}},
  howpublished = {GitHub repository},
  url = {https://github.com/yym68686/uni-api},
  year = {2026},
  note = {Evaluated commit 20da7a7cc6f71b9a2dbb42ff215c2cbda6ce8e43 (2026-07-19)},
}

@misc{metapi_repo,
  author = {{MetaAPI Contributors}},
  title = {{MetaAPI}},
  howpublished = {GitHub repository},
  url = {https://github.com/cita-777/metapi},
  year = {2026},
  note = {Evaluated commit 41767a65ec8e5470a9a70f4615b47dc24949afff (2026-06-24)},
}

@misc{litellm_repo,
  author = {{BerriAI Contributors}},
  title = {{LiteLLM}},
  howpublished = {GitHub repository},
  url = {https://github.com/BerriAI/litellm},
  year = {2026},
  note = {Evaluated commit 7e66f00fca08550811e6e73d18de25cd90ab7221 (2026-07-21)},
}

@misc{sub2api_repo,
  author = {{Sub2API Contributors}},
  title = {{Sub2API}},
  howpublished = {GitHub repository},
  url = {https://github.com/Wei-Shaw/sub2api},
  year = {2026},
  note = {Evaluated commit b8b72e1b18310c908668e79112e43c1e7c682696 (2026-07-21)},
}

@misc{oneapi_repo,
  author = {{Songquanpeng Contributors}},
  title = {{One API}},
  howpublished = {GitHub repository},
  url = {https://github.com/songquanpeng/one-api},
  year = {2026},
  note = {Frozen lineage baseline; accessed 2026-08-02},
}

@misc{openrouter_about,
  author = {{OpenRouter}},
  title = {{About OpenRouter}},
  year = {2026},
  url = {https://openrouter.ai/about},
  note = {Snapshot accessed 2026-08-02},
}

@misc{borri2026aipremium,
      title={AI Premium}, 
      author={Nicola Borri and Yukun Liu and Aleh Tsyvinski},
      year={2026},
      eprint={2606.30583},
      archivePrefix={arXiv},
      primaryClass={cs.CY},
      url={https://arxiv.org/abs/2606.30583}, 
}

@misc{ramp_openrouter,
  author = {{Ramp}},
  title = {{OpenRouter Ramp Rate: A Data-Backed Look}},
  year = {2026},
  url = {https://ramp.com/vendors/openrouter},
  note = {Transaction-panel definitions and snapshot accessed 2026-08-02},
}

@techreport{nist_deepseek,
  author = {{Center for AI Standards and Innovation}},
  title = {{Evaluation of DeepSeek AI Models}},
  institution = {National Institute of Standards and Technology},
  year = {2025},
  month = sep,
  url = {https://www.nist.gov/system/files/documents/2025/09/30/CAISI_Evaluation_of_DeepSeek_AI_Models.pdf},
}

@misc{openai_prompt_cache,
  author = {{OpenAI}},
  title = {{Prompt Caching}},
  year = {2026},
  url = {https://developers.openai.com/api/docs/guides/prompt-caching},
  note = {Documentation snapshot accessed 2026-08-02},
}

@misc{openai_projects,
  author = {{OpenAI}},
  title = {{Managing Projects in the API Platform}},
  year = {2026},
  url = {https://help.openai.com/en/articles/9186755-managing-projects-in-the-api-platform},
  note = {Documentation snapshot accessed 2026-08-02},
}

@misc{anthropic_workspaces,
  author = {{Anthropic}},
  title = {{Workspaces}},
  year = {2026},
  url = {https://platform.claude.com/docs/en/manage-claude/workspaces},
  note = {Documentation snapshot accessed 2026-08-02},
}

@misc{anthropic_cache,
  author = {{Anthropic}},
  title = {{Prompt Caching}},
  year = {2026},
  url = {https://platform.claude.com/docs/en/build-with-claude/prompt-caching},
  note = {Documentation snapshot accessed 2026-08-02},
}

@misc{google_gemini_cache,
  author = {{Google}},
  title = {{Context Caching}},
  year = {2026},
  url = {https://ai.google.dev/gemini-api/docs/generate-content/caching},
  note = {Gemini API documentation snapshot accessed 2026-08-04},
}

@misc{google_gemini_pricing,
  author = {{Google}},
  title = {{Gemini Developer API Pricing}},
  year = {2026},
  url = {https://ai.google.dev/gemini-api/docs/pricing},
  note = {Context-cache storage pricing snapshot accessed 2026-08-04},
}

@misc{openai_second_org,
  author = {{OpenAI}},
  title = {{Can I Create an Additional Platform API Organization?}},
  year = {2026},
  url = {https://help.openai.com/en/articles/8991840-can-i-create-an-additional-platform-api-organization},
  note = {Help Center article; states an additional API organization is created by OpenAI on request rather than self-serve. Retrieved 2026-08-08},
}

@misc{openai_services_agreement,
  author = {{OpenAI}},
  title = {{Service Terms and Business Terms}},
  year = {2026},
  url = {https://openai.com/policies/services-agreement/},
  note = {Prohibits sharing, renting, or reselling accounts or seats and selling, buying, or transferring API keys, while permitting a customer to build products on the API for its own end users; enterprise agreements may vary. Retrieved 2026-08-08},
}

@misc{anthropic_rate_limits,
  author = {{Anthropic}},
  title = {{Rate Limits}},
  year = {2026},
  url = {https://platform.claude.com/docs/en/api/rate-limits},
  note = {Documentation snapshot retrieved 2026-08-08},
}

@misc{anthropic_commercial_terms,
  author = {{Anthropic}},
  title = {{Commercial Terms of Service}},
  year = {2026},
  url = {https://www.anthropic.com/legal/commercial-terms},
  note = {Section D.4 restricts reselling the Services except as expressly approved. Retrieved 2026-08-08},
}

@ARTICLE{song2025earlybird,
  author={Song, Linke and Pang, Zixuan and Wang, Wenhao and Wang, Zihao and Wang, XiaoFeng and Chen, Hongbo and Song, Wei and Jin, Yier and Meng, Dan and Hou, Rui},
  journal={IEEE Transactions on Information Forensics and Security}, 
  title={The Early Bird Catches the Leak: Unveiling Timing Side Channels in LLM Serving Systems}, 
  year={2025},
  volume={20},
  number={},
  pages={11431-11446},
  doi={10.1109/TIFS.2025.3622954}}

@InProceedings{gu2025auditing,
  title = 	 {Auditing Prompt Caching in Language Model {API}s},
  author =       {Gu, Chenchen and Li, Xiang Lisa and Kuditipudi, Rohith and Liang, Percy and Hashimoto, Tatsunori},
  booktitle = 	 {Proceedings of the 42nd International Conference on Machine Learning},
  pages = 	 {20477--20496},
  year = 	 {2025},
  editor = 	 {Singh, Aarti and Fazel, Maryam and Hsu, Daniel and Lacoste-Julien, Simon and Berkenkamp, Felix and Maharaj, Tegan and Wagstaff, Kiri and Zhu, Jerry},
  volume = 	 {267},
  series = 	 {Proceedings of Machine Learning Research},
  month = 	 {13--19 Jul},
  publisher =    {PMLR},
  url = 	 {https://proceedings.mlr.press/v267/gu25b.html}
}

@inproceedings{wu2025promptpeek,
  author = {Guanlong Wu and Zheng Zhang and Yao Zhang and Weili Wang and Jianyu Niu and Ye Wu and Yinqian Zhang},
  title = {{I Know What You Asked: Prompt Leakage via KV-Cache Sharing in Multi-Tenant LLM Serving}},
  year = {2025},
  booktitle = {Network and Distributed System Security (NDSS) Symposium},
  doi = {10.14722/ndss.2025.241772},
  url = {https://doi.org/10.14722/ndss.2025.241772},
}

@misc{zheng2024inputsnatch,
      title={InputSnatch: Stealing Input in LLM Services via Timing Side-Channel Attacks}, 
      author={Xinyao Zheng and Husheng Han and Shangyi Shi and Qiyan Fang and Zidong Du and Xing Hu and Qi Guo},
      year={2024},
      eprint={2411.18191},
      archivePrefix={arXiv},
      primaryClass={cs.CR},
      url={https://arxiv.org/abs/2411.18191}, 
}

@misc{fahey2026cacheprobe,
      title={CacheProbe: Auditing Prompt Cache Isolation in Gateway APIs}, 
      author={Ryan Fahey},
      year={2026},
      eprint={2605.30613},
      archivePrefix={arXiv},
      primaryClass={cs.CR},
      url={https://arxiv.org/abs/2605.30613}, 
}

@misc{wang2026optileak,
      title={OptiLeak: Efficient Prompt Reconstruction via Reinforcement Learning in Multi-tenant LLM Services}, 
      author={Longxiang Wang and Xiang Zheng and Xuhao Zhang and Yao Zhang and Ye Wu and Cong Wang},
      year={2026},
      eprint={2602.20595},
      archivePrefix={arXiv},
      primaryClass={cs.CR},
      url={https://arxiv.org/abs/2602.20595}, 
}

@misc{sun2026spliceleak,
      title={Agent-Assisted Side-Channel Attacks on Non-Prefix KV Cache in RAG}, 
      author={He Sun and Shinan Liu and Siyuan Ma and Junhao Li and Mingjun Xiao and Wenhao Jiang},
      year={2026},
      eprint={2606.21842},
      archivePrefix={arXiv},
      primaryClass={cs.CR},
      url={https://arxiv.org/abs/2606.21842}, 
}

@misc{chu2025safekv,
      title={Selective KV-Cache Sharing to Mitigate Timing Side-Channels in LLM Inference}, 
      author={Kexin Chu and Zecheng Lin and Dawei Xiang and Zixu Shen and Jianchang Su and Cheng Chu and Yiwei Yang and Wenhui Zhang and Wenfei Wu and Wei Zhang},
      year={2026},
      eprint={2508.08438},
      archivePrefix={arXiv},
      primaryClass={cs.CR},
      url={https://arxiv.org/abs/2508.08438}, 
}

@misc{wu2026cacheprune,
      title={CachePrune: Privacy-Aware and Fine-Grained KV Cache Sharing for Efficient LLM Inference}, 
      author={Guanlong Wu and Zhaohan li and Yao Zhang and Zheng Zhang and Jianyu Niu and Ye Wu and Yinqian Zhang},
      year={2026},
      eprint={2605.23640},
      archivePrefix={arXiv},
      primaryClass={cs.CR},
      url={https://arxiv.org/abs/2605.23640}, 
}

@misc{pennas2026prefixwall,
      title={PrefixWall: Mitigating Prefix Caching Side Channels in Shared LLM Systems}, 
      author={Panagiotis Georgios Pennas and Konstantinos Papaioannou and Marco Guarnieri and Thaleia Dimitra Doudali},
      year={2026},
      eprint={2603.10726},
      archivePrefix={arXiv},
      primaryClass={cs.CR},
      url={https://arxiv.org/abs/2603.10726}, 
}

@inproceedings {mirheidari2020cachedconfused,
author = {Seyed Ali Mirheidari and Sajjad Arshad and Kaan Onarlioglu and Bruno Crispo and Engin Kirda and William Robertson},
title = {Cached and Confused: Web Cache Deception in the Wild},
booktitle = {29th USENIX Security Symposium (USENIX Security 20)},
year = {2020},
isbn = {978-1-939133-17-5},
pages = {665--682},
url = {https://www.usenix.org/conference/usenixsecurity20/presentation/mirheidari},
publisher = {USENIX Association},
month = aug
}

@INPROCEEDINGS{bosman2016dedup,
  author={Bosman, Erik and Razavi, Kaveh and Bos, Herbert and Giuffrida, Cristiano},
  booktitle={2016 IEEE Symposium on Security and Privacy (SP)}, 
  title={Dedup Est Machina: Memory Deduplication as an Advanced Exploitation Vector}, 
  year={2016},
  volume={},
  number={},
  pages={987-1004},
  doi={10.1109/SP.2016.63}}

@misc{vllm_cache_salt,
  author = {{vLLM Project}},
  title = {{Automatic Prefix Caching}},
  year = {2026},
  howpublished = {vLLM documentation},
  url = {https://docs.vllm.ai/en/latest/design/prefix_caching/},
  note = {See ``Cache Isolation for Security''; accessed 2026-08-03},
}

@misc{sglang_radix,
  author = {{SGLang Project}},
  title = {{Server Arguments}},
  year = {2026},
  howpublished = {SGLang documentation},
  url = {https://github.com/sgl-project/sglang/tree/main},
  note = {See the RadixAttention prefix-cache option; accessed 2026-08-03},
}

\appendices
\setcounter{topnumber}{3}
\setcounter{totalnumber}{4}
\setcounter{dbltopnumber}{3}
\renewcommand{\topfraction}{0.92}
\renewcommand{\dbltopfraction}{0.95}
\renewcommand{\textfraction}{0.06}
\renewcommand{\floatpagefraction}{0.70}
\renewcommand{\dblfloatpagefraction}{0.70}
\makeatletter
\setlength{\@fptop}{0pt}
\setlength{\@fpsep}{16pt}
\setlength{\@fpbot}{0pt plus 1fil}
\setlength{\@dblfptop}{0pt}
\setlength{\@dblfpsep}{16pt}
\setlength{\@dblfpbot}{0pt plus 1fil}
\makeatother
\section{Model Properties and Supporting Details}
\label{app:formal-results}

This section states properties of the measured observable that support the main methodology and defense.  We first formalize directed reachability, then show why cache writes by an attacker obscure provenance.  We next give the request complexity and conditions for sequential composition, followed by the effects of reporting granularity and route instability.  The final two subsections state conditional separation at the final boundary and quantify structural colocation when enforced domains are scarce.

\subsection{Directed reachability and its symmetric special case}

For a fixed history $H$, time $t$, and observable, request $q$ reaches $q'$ if executing $q$ changes the later transcript for $q'$ through cache state.  The relation is generally directed.  Suppose lookup and write selection is deterministic, remains independent of requests within a domain, is mutually symmetric, and remains stable during residence, while the observable depends only on shared domain state.  Under these premises, mutual reachability is reflexive, symmetric, and transitive and therefore induces equivalence classes. Experiments through a black box establish individual directed edges without assuming these premises globally.

\subsection{Provenance under cache writes}

The directed edge above leaves open whether victim state or a later attacker write caused an observation, so we now formalize provenance.  Let $R(q,S)$ be deterministic and monotone under inclusion of entry sets, and let $V$ and $A$ be victim and attacker histories.  For an observable based on the maximum matching prefix, if $R(q,A)\ge R(q,V)$, then $R(q,V\cup A)=R(q,A)$; removing the victim leaves the observation unchanged.  This proves why input uniqueness and a later owner hot repeat cannot identify the victim.

Attribution therefore needs a reference that isolates the victim contribution.  Consider a paired shadow $A'$ whose attacker history contains the identical tokens with no victim state.  When probe and shadow share route, formation, expiry, eviction, tokenizer, quantizer, prompt length, and noise, and the floors for the known prefix are comparable, then
\[
 R(q,V\cup A)>R(q,A')
 \quad\Longleftrightarrow\quad R(q,V)>R(q,A),
\]
by substituting $R(q,A')=R(q,A)$ and expanding the primary maximum.  In this comparable case the probe reaches no deeper than the victim prefix, so a victim contribution appears as an absolute excess over the shadow.

A live probe need not meet the comparable floor premise.  When the attacker's probe is longer than the victim's stored prefix, as in the Section~\ref{sec:extraction} chain where the victim wrote a known context and a short secret, the primary reads the victim prefix while the shadow reads the attacker's own near full match, so the two floors differ by a fixed prompt length offset.  Provenance then uses the relative delta of Equation~\ref{eq:relative-selector}: every candidate is scored against the same shadow reference, a correct extension advances the match one token past the shared baseline, and a complete replay with no victim must return every candidate to that baseline.  A lift present with victim state and absent in the replay is attributable to the victim whatever the absolute floor.  Asymmetric error, retry, route, or formation breaks the required equalities in either case.

\subsection{Sequential complexity and conditional composition}

We next state when such decisions reduce sequential search.  Given an exact attributable extension oracle and candidate set $C_i$ at step $i$, sequential recovery tests at most $\sum_i|C_i|$ candidates: test candidates until one is accepted, commit it, and continue.  Testing complete strings over the Cartesian product may require $\prod_i|C_i|$.  These expressions count logical candidate decisions; gate, shadow, and replay requests add bounded factors.  Elimination can save the final test, so the sum is a conservative upper bound.

Decisions supplied with the correct previous tokens compose into a stateful path when four premises hold: (i) every correct commit leaves a raw prefix and attack state observationally identical to the next fixture's initial state; (ii) the oracle adds no unmodeled state across positions, or an equivalent reset is used; (iii) routing, formation, provenance, and residence persist; and (iv) total work fits the path budget.  The 192 BIP-39 positions evaluate individual positions.  When all four premises hold, the decisions compose without an additional failure of cache isolation.

\subsection{Granularity, alignment, and budget}

Sequential composition depends on whether reporting distinguishes each correct extension, so we now make granularity explicit.  For downward quantization $Q_g(z)=g\lfloor z/g\rfloor$ and verified prefix length $L$, the next boundary is $\Delta(L)=g-(L\bmod g)$, interpreting zero as $g$.  A correct token followed by $s$ known matching tokens becomes visible exactly when $1+s\ge \Delta(L)$.  Thus $g=1$ exposes each correct next token, while alignment or known suffixes can defeat coarse reporting.  With no known suffix, a position is visible exactly when $L\equiv g-1 \pmod g$, so a set of $n$ positions whose verified lengths were uniform modulo $g$ would expose $n/g$ of them in expectation.  That expectation is not a bound: an empirical set moves above or below it as its lengths depart from uniform.

Let $\mathrm{rk}_i$ be the first rank of a raw candidate that crosses the hidden boundary.  Under an exact oracle, delayed commit, and sufficient residence, an $n$ step path succeeds exactly when $\mathrm{rk}_i\le B_i$ for every budget $B_i$ at step $i$.  If every reachable history has pointwise sequential conditional minimum entropy $\eta_i$, at most $B_i$ candidates plus one final guess give
\[
 \Pr[\text{recover step }i]\le
 \min\{1,(B_i+1)2^{-\eta_i}\}.
\]
Average surprisal from a language model does not provide this pointwise lower bound.

\subsection{Nonimplications across route and time}

We next show why an isolated miss establishes neither a comparable route nor a resident entry.  With residence $E(t)$ and route availability $U(t)$, $Y(t)=E(t)U(t)$ implies that a hit places a lower bound on residence while a miss provides no upper bound.  Likewise, jointly hot primary and shadow frontiers need not imply a symmetric candidate lift: routing based on the complete request or sharding based on the prefix supplies a counterexample.  The measured Relay N cell instantiates this result with six jointly hot frontiers and 0/2 symmetric first extensions.

\subsection{Conditional separation at the final boundary}

These route and time effects motivate a defense stated over every final operation.  Assume the relay derives $n_d$ from authenticated customer $d$; clients cannot forge it; every final lookup and write includes it; no weaker fallback exists; and the provider enforces it, a primitive no provider we measured currently exposes.  Conditioned on absence of an active collision, entries written under $d$ lie outside every lookup and write domain for $d'\ne d$, eliminating both directions of cache reachability while preserving reuse by the same customer.  For pseudorandom function (PRF) outputs of $k$ bits, which the HMAC construction supplies, $N$ active customers, and $w$ accepted versions per customer, the union bound on failure is $(wN)^2/2^{k+1}$, which the reference parameter $k=128$ of Section~\ref{sec:defenses} keeps negligible for any deployed population.  Dropping the namespace on any final path supplies a direct counterexample.  This argument establishes separation of cache state; full transcript isolation additionally covers billing, timing, compute, errors, routes, quotas, and future state.

Binding the model $m$ and the provider or deployment $v$ into the namespace is a conservative policy choice rather than a requirement of the theorem.  It prevents reuse across deployments that differ in tokenization, cache semantics, or security domain.  An operator may omit $m$ when the final provider documents namespace semantics across models and the resulting reuse stays inside one security domain.

\subsection{Capacity of enforced domains}

We conclude by quantifying unavoidable colocation when enforced domains are scarce.  Following the notation of Section~\ref{sec:identity-theory}, map $N$ downstream domains into $D\ge1$ domains enforced by the provider.  Convexity of $\binom{x}{2}$ implies that an allocation minimizing pairs differs in bin size by at most one.  With $a=\lfloor N/D\rfloor$ and $b=N\bmod D$, it has $b$ bins of size $a+1$ and $D-b$ of size $a$, yielding $\Pi(N,D)=b\binom{a+1}{2}+(D-b)\binom{a}{2}$ colocated pairs.  Dividing by $\binom{N}{2}$ gives a structural collision bound for a uniformly selected pair under the fixed mapping.  Actual \aone{} reachability additionally requires cache formation, a compatible route, a matching prefix, residence, and an observable.

\section{Extended Methodology and Results}
\label{app:extended-methodology}

This appendix supplies the detailed configurations, denominators, and evidence ledgers behind the main results.  We first record the frozen gateway configurations, then the documented supply of cache domains for each credential class and the credential selection each project ships by default.  We next qualify the explicit cache ingress, describe production classification and the OpenRouter cohort, and expand the defense cells and the ecosystem denominators.  The remaining subsections document extraction implementation and accounting, local candidate studies, eligibility rules, abstaining and historical records, and timing and reproducibility.

For evidence bookkeeping, \evidence{E0} denotes a formal or model claim; \evidence{E1} denotes source analysis, local execution, simulation, or corpus replay; \evidence{E2} denotes a gateway configured by the researchers and executed against a real upstream model with closed weights; and \evidence{E3} denotes a provider or production relay measurement using accounts controlled by the researchers.  The labels identify the execution environment.  Causal strength additionally depends on whether the design assigns an intervention or observes an existing path.

\begin{table*}[t]
\centering
\caption{Measurement contract. Each row names an experimental unit, its controls, and the strongest conclusion supported by a valid result. Repeated requests within one cell serve as controls.}
\label{tab:measurement-contract}
\small
\setlength{\tabcolsep}{4pt}
\renewcommand{\arraystretch}{1.12}
\begin{tabularx}{\textwidth}{@{}L{0.50}L{0.85}L{1.80}L{0.85}@{}}
\toprule
Question & Experimental unit & Design and validity controls & Permitted inference \\
\midrule
\textbf{RQ1}\newline Existence & Model and provider route $\times$ account pair $\times$ dated window & Four request unit; fresh prefix; formation control through an owner repeat; unrelated fresh negative; comparable model, route, status, and retry behavior & A directed cache edge between customers on that path and window \\
\textbf{RQ2}\newline Localization & Matched path or intervention arms & Principal and namespace splits; pool partition recorded by observer; adapter contrast; nested overwrite versus preservation & Cause at the component level for an assigned intervention, selector association for an observed choice, or bundled contrast as declared \\
\textbf{RQ3}\newline Capability & One frozen dependent chain, or the declared corpus or fixture unit & Primary and shadow histories; preheating that accounts for cache writes; route and formation gates; complete replay without victim state; abstention when a gate fails & Composition attributable to victim state on an eligible path, or a measured limit for one component \\
\textbf{RQ4}\newline Defense & Gateway and path $\times$ defense arm & Shared, forged client, namespace derived by relay, and separate principal arms; reuse by the same customer and controls for failure paths & Conditional separation of cache state on tested paths \\
\bottomrule
\end{tabularx}
\end{table*}

\subsection{Frozen gateway configurations}

The E2 study used NewAPI commit \texttt{e0d515611588}, uni-api \texttt{20da7a7cc6f7}, MetaAPI \texttt{41767a65ec8e}, LiteLLM \texttt{7e66f00fca08}, and Sub2API \texttt{b8b72e1b1831}.  One API is the stale baseline used only for lineage.  Each executable gateway ran locally with two authenticated downstream customers.  The observer forwarded real OpenAI Chat Completions and native Anthropic Messages calls, recorded digests of requests and authorization together with status, model, and usage metadata, and returned the provider response.  Formal cells with four requests allowed no retries and required one completion token, a cold prime and negative, and a valid later owner hot repeat.  Diagnostics in which a prime or negative was already hot, or which contained extra calls or route errors, were excluded before the frozen run.  Across the five projects, 188 interface tests spanning source and runtime verified data flow before the provider cache experiments.

Table~\ref{tab:configuration-realism} records the supported configuration represented by each cell.  The frozen documentation at the top level describes authentication, quota, routing, and load balancing.  It contains no explicit promise that each downstream key obtains its own enforced domain.

\begin{table*}[t]
\centering
\caption{Configuration realism and identity behavior of individual adapters at the frozen revisions.  ``Instantiated'' means that we assigned two controlled downstream customers to one controlled upstream principal using documented project objects, which is the mapping the domain supply of Appendix~\ref{app:credential-classes} leaves available at scale.}
\label{tab:configuration-realism}
\small
\setlength{\tabcolsep}{3.2pt}
\renewcommand{\arraystretch}{1.14}
\begin{tabularx}{\textwidth}{@{}L{0.42}L{1.00}L{1.68}L{0.90}@{}}
\toprule
Project & Documented mapping & Frozen path behavior & Evidence label and design \\
\midrule
NewAPI & Downstream token or user to channel and key pool & Generic path selected the shared provider credential and preserved a caller cache key. & Instantiated E2 \\
uni-api & Downstream API keys to provider and key pool with load balancing & Provider selection omitted downstream customer identity. & Instantiated E2 \\
MetaAPI & Managed key and policy to route and site token pool & Upstream site token replaced downstream authentication; local stickiness did not select a provider namespace. & Instantiated E2 \\
LiteLLM & Virtual key, team, or organization to shared deployment & Tested Chat and Responses surfaces both dropped the caller field and shared cache state. & Instantiated E2 \\
Sub2API & Downstream key and group to account pool & Generic OpenAI forwarding preserved the caller field; Grok adapter instead derived a namespace from the downstream key and model. & Instantiated E2 plus control adapter \\
One API & User or token group to channels and load balancing & Stale NewAPI lineage supplied source evidence. & Source and lineage only \\
\bottomrule
\end{tabularx}
\end{table*}

The same audit constructed a graph with an exclusive upstream only after confirming
multiple downstream principals.  NewAPI uses a unique group and a channel with one key;
uni-api uses a rule for each key qualified by provider; MetaAPI uses an
authoritative \texttt{allowedRouteIds} policy and a route with one channel; LiteLLM
uses a BYOK deployment scoped to one team; and Sub2API uses an account that appears in
exactly one exclusive group.  Each construction also removes shared retry and
fallback targets.  These are gateway routing preconditions.  They become
cache security boundaries when the selected credentials belong to distinct
domains enforced by the provider.

\subsection{Credential classes and domain supply}
\label{app:credential-classes}

Section~\ref{sec:identity-theory} defines the security quantity as $D$, the enforced domains a path can obtain.  This appendix records the documented supply of $D$ for each credential class a relay can hold, and the published evidence that one class exposes no isolation primitive at all.  It establishes supply and source; it does not estimate how many deployments pool.  Table~\ref{tab:credential-classes} summarizes the classes, and Table~\ref{tab:defense-guarantees} records the tier that Section~\ref{sec:defenses} assigns to each deployment primitive.

\begin{table}[t]
\centering
\caption{Deployment primitive and defensible claim.}
\label{tab:defense-guarantees}
\small
\setlength{\tabcolsep}{3.5pt}
\renewcommand{\arraystretch}{1.12}
\begin{tabularx}{\columnwidth}{@{}L{0.80}L{1.20}@{}}
\toprule
Primitive & Defensible claim \\
\midrule
Separate provider principal/domain & Cache boundary documented by the provider; preferred root defense. \\
Opaque namespace enforced by provider & Conditional separation of cache state if every path propagates the namespace. \\
Routing hint & Measured lookup mitigation only; semantics may change. \\
Visible marker, coarsening, hiding & Controls attack cost. \\
\bottomrule
\end{tabularx}
\end{table}

\begin{table*}[t]
\centering
\caption{Documented cache domain supply and terms status by upstream credential class.  Vendor documentation retrieved 2026-08-08; gateway evidence is from the frozen revisions.}
\label{tab:credential-classes}
\small
\setlength{\tabcolsep}{3.2pt}
\renewcommand{\arraystretch}{1.14}
\begin{tabularx}{\textwidth}{@{}L{1.15}L{0.62}L{1.60}L{0.63}@{}}
\toprule
Credential class & Finest documented cache domain & Supply of $D$ & Terms status \\
\midrule
OpenAI platform key & Organization & Additional organization provisioned on request, not self-serve & Within terms \\
Anthropic key, direct API and two partner platforms & Workspace & 100 workspaces per organization, sharing one rate limit & Within terms \\
Anthropic key via Amazon Bedrock or Google Cloud & Organization & One per organization & Within terms \\
Consumer subscription OAuth or setup token & None exposed & One per account & Outside terms \\
\bottomrule
\end{tabularx}
\end{table*}

On OpenAI, prompt caches are not shared between organizations and only members of one organization reach a cache for identical prompts; projects select data controls but no separate cache scope, and an additional organization is provisioned by the vendor on request rather than created by the customer~\cite{openai_prompt_cache,openai_projects,openai_second_org}.  The documented request field for cache behavior is described as influencing routing and hit rates rather than as an access control, which is the evidence for placing it in the routing tier of Table~\ref{tab:defense-guarantees}.

On Anthropic, caches are isolated per workspace on the direct API and two partner platforms and per organization on Amazon Bedrock and Google Cloud~\cite{anthropic_cache}.  Workspaces are creatable through the administrative API, so the boundary can be automated, but one organization admits at most 100 of them, and organization limits continue to bound their aggregate because workspace limits partition the organization's allowance instead of adding to it~\cite{anthropic_workspaces,anthropic_rate_limits}.  The same rate limit documentation excludes cache reads from the input token limit, which is why separating every customer costs throughput as well as price.

The terms determine which classes a compliant operator may hold.  OpenAI's services agreement prohibits sharing, renting, or reselling accounts or seats and prohibits selling, buying, or transferring API keys, while permitting a customer to build a product on the API and offer that product to its own end users; individual enterprise agreements may vary these defaults~\cite{openai_services_agreement}.  Anthropic's commercial terms restrict a customer from reselling the services except as expressly approved~\cite{anthropic_commercial_terms}.  A relay that purchases platform capacity and resells inference to its own customers is therefore ordinarily within terms, whereas a deployment that redistributes seats or subscription quota is not.

Evidence for the subscription class comes from published source at the frozen revisions rather than from probing.  Sub2API describes itself as a gateway platform for subscription quota distribution and warns in its own documentation that use may violate the terms of upstream providers~\cite{sub2api_repo}.  Its account model admits an OAuth type carrying profile and inference scope and a setup token type carrying inference scope, alongside ordinary key and upstream types, and its authorization endpoints address the consumer surfaces of Anthropic, OpenAI, and xAI rather than the platform APIs.  NewAPI carries a coding agent channel whose credential is an OAuth refresh token~\cite{newapi_repo}.  Credentials of this class expose no workspace, project, or administrative interface, so $D=1$ per account and colocation is not a setting an operator can change.  This class also has no disclosure counterparty, which constrains the remediation path recorded in the ethics section.

\subsection{Default credential selection posture}
\label{app:default-posture}

Section~\ref{sec:causal-evaluation} executes an exclusive credential graph that an operator must author.  This appendix records what each project does before that authoring, because a default that already pools makes the configured mapping a reproduction of the shipped posture rather than a departure from it.  We define the default as the shipped configuration of the frozen revision plus only the actions needed to make the gateway serve two customers, namely creating the customers and registering credentials; where a project ships a quickstart, its documented minimal configuration counts as default.

The determination is source level.  The selection policy is a property of the code and its shipped defaults, whereas a runtime trace samples one path through that policy, so source is the stronger evidence and uses the same tracing stage as the main method.  Each finding is a probe declaring a file, a pattern, and whether the pattern must be present or absent; a probe that stops matching is reported as broken rather than dropped.  Every negative probe also names a sentinel known to occupy the same window, and a probe whose sentinel stops matching is reported as broken instead of confirming, so an absent verdict cannot pass on an empty or relocated window.  All 21 probes confirm, and the sentinel is present for all three negative probes.  No gateway was executed and no provider was contacted.

\begin{table*}[t]
\centering
\caption{Shipped default credential selection at the frozen revisions.  No project binds a downstream customer to an upstream credential before an operator authors that binding.}
\label{tab:default-posture}
\small
\setlength{\tabcolsep}{3.2pt}
\renewcommand{\arraystretch}{1.14}
\begin{tabularx}{\textwidth}{@{}L{0.40}L{0.62}L{0.80}L{2.18}@{}}
\toprule
Gateway & Revision & Default selection & Mechanism establishing the default \\
\midrule
NewAPI & \texttt{e0d515611588} & Uniform random per request & Selection takes group and model with no customer parameter; channels default to weight and priority zero, which the selector rewrites to equal effective weight before a random draw \\
uni-api & \texttt{20da7a7cc6f7} & Round robin & The provider key pool defaults to round robin, and the quickstart grants one downstream key every model in every channel \\
MetaAPI & \texttt{41767a65ec8e} & Weighted random, fail open (permits by default) & The route strategy defaults to weighted, and a key with no model patterns and no allowed routes returns permitted because the shipped empty policy omits the deny flag \\
LiteLLM & \texttt{7e66f00fca08} & Simple shuffle & The router strategy defaults to simple shuffle, and any model not claimed by an explicit routing group falls into one implicit group \\
Sub2API & \texttt{b8b72e1b1831} & Seeded, then randomized & The seed uses session, response, model, and group anchors but no customer, and an anchorless request receives time entropy so that it does not stay on one account \\
\bottomrule
\end{tabularx}
\end{table*}

Table~\ref{tab:default-posture} summarizes the outcome.  The result bounds what the configured experiment can be accused of: because no shipped default binds a customer to a credential, an operator who buys separate credentials in order to separate customers still receives none, and three of the projects actively redistribute a single customer across the credential set.  Two non-default actions are therefore required for separation, namely authoring one binding object per customer and suppressing cross-credential fallback.  The appendix establishes the shipped default and the mechanism behind it.  It does not estimate how many operators pool in practice, and it does not claim that binding is unavailable; the main text records the binding object each project offers.

\subsection{Qualification of explicit cache ingress}

\begin{table*}[t]
\centering
\caption{Provider cache formation and the interface delegated by evaluated relays.  Gemini cells sent a real create method, and the successful generic interface terminated at a local fixture.}
\label{tab:provider-compatibility}
\footnotesize
\setlength{\tabcolsep}{3.5pt}
\renewcommand{\arraystretch}{1.10}
\begin{tabularx}{\textwidth}{@{}L{0.62}L{0.95}L{0.78}L{1.35}L{1.30}@{}}
\toprule
Surface & Native formation & Documented domain & Evaluated relay behavior & Security inference \\
\midrule
OpenAI automatic & Eligible inference request & Organization & Supported through ordinary inference; reads between customers on tested compatible paths & Shared upstream identity can immediately expose state. \\
Anthropic opt in & Request \texttt{cache\_control} & Workspace or upstream organization & Caching enabled by the request incurred a creation charge and allowed a later read between customers & The opt in requirement limits affected requests. \\
Gemini implicit & Inference cache managed by provider & Project & Ordinary inference supported; valid tested cells reported no read between customers & Dated empirical result for the tested paths. \\
Gemini explicit & Create \texttt{CachedContent}, then reference its handle & Project & Four ingresses rejected create; LiteLLM required an explicit route grant to forward creation and reference & Lifecycle availability depends on adapter and authorization; the success used a local fixture. \\
\bottomrule
\end{tabularx}
\end{table*}

The frozen inference configurations above define ordinary model paths; Gemini's explicit cache requires a separate control interface that we qualify next.  Every Gemini cell for this control interface submitted an actual cache creation request.  Table~\ref{tab:gemini-create-ingress}
records the first response and whether it reached the observer.  A reference was sent only after creation returned a resource name.
The successful LiteLLM generic interface supplies gateway protocol evidence through the local fixture.

\begin{table}[t]
\centering
\caption{Actual Gemini explicit cache creation attempts at frozen gateway ingresses.}
\label{tab:gemini-create-ingress}
\small
\setlength{\tabcolsep}{4pt}
\renewcommand{\arraystretch}{1.12}
\begin{tabular*}{\columnwidth}{@{\extracolsep{\fill}}lccc@{}}
\toprule
Gateway/arm & Create & Reached observer & Reference \\
\midrule
NewAPI & 404 & No & N/A \\
uni-api & 405 & No & N/A \\
MetaAPI & 404 & No & N/A \\
LiteLLM default & 403 & No & N/A \\
LiteLLM granted & 200 & Yes & 200 \\
Sub2API & 404 & No & N/A \\
\bottomrule
\end{tabular*}
\end{table}

The cached input count was calibrated separately from these E2 cells. Table~\ref{tab:granularity-calibration} traces six requests on one dated \texttt{grok-4-1-fast-non-reasoning} production route. The initial target and fresh negative reported 162 and 163 cached tokens, which establishes the background reuse visible on this route, and the exact repeat rose to 768. Mutating the word at positions 128, 129, and 130 then yielded 482, 483, and 484 cached tokens. The reported values are provider token counts, so three unit steps in word position produced three unit steps in the reported count, which establishes one token resolution in the provider field even though a fresh request already includes background reuse. A later preregistered replication of this cell aborted with the same prefix cache control too small to enter candidates, so we report the calibration as a route specific observation rather than a stable oracle.

\begin{table}[t]
\centering
\caption{Production reporter calibration at a fixed prompt family.}
\label{tab:granularity-calibration}
\small
\setlength{\tabcolsep}{5pt}
\renewcommand{\arraystretch}{1.13}
\begin{tabular*}{\columnwidth}{@{\extracolsep{\fill}}lrr@{}}
\toprule
Request & Prompt tokens & Cached tokens \\
\midrule
Initial target & 769 & 162 \\
Exact repeat & 769 & 768 \\
Mutation at word 128 & 769 & 482 \\
Mutation at word 129 & 769 & 483 \\
Mutation at word 130 & 769 & 484 \\
Fresh negative & 770 & 163 \\
\bottomrule
\end{tabular*}
\end{table}

A single route does not fix the granularity for a model family.  We submitted the same six request calibration to eleven Grok labels the relay exposed on one dated frame.  Table~\ref{tab:grok-family-granularity} reports the adjacent mutation reads and the resulting classification.  A label is strict one token when all three reads rise by one, one token on a hot path when that rise appears only once the path is already hot, and coarse, two token, gapped, or unstable otherwise.  One label reported strict one token resolution across all three adjacent mutations, four exposed a one token step only where the path was already hot, two reported a coarse 128 token block with no one token step, and the remaining four were gapped, two token, unstable, or incomplete.  No single quantum describes the family, and the served label \texttt{grok-4-fast-non-reasoning} used in the Section~\ref{sec:extraction} chain fell in the hot path class that the chain's gates enforce.  Reporting granularity, like the identity boundary, is therefore a property of the concrete route rather than of the model label.

\begin{table*}[t]
\centering
\caption{Reporting granularity across eleven Grok labels on one relay and dated frame.  Adjacent mutation reads are the cached token counts at three consecutive word positions; the same calibration and prompt family are used for every label.}
\label{tab:grok-family-granularity}
\footnotesize
\setlength{\tabcolsep}{5pt}
\renewcommand{\arraystretch}{1.12}
\begin{tabular*}{\textwidth}{@{\extracolsep{\fill}}lrll@{}}
\toprule
Model label & Repeat cached & Adjacent mutation reads & Classification \\
\midrule
grok-4-1-fast-non-reasoning & 650 & 484, 485, 486 & Strict one token \\
grok-4-fast-non-reasoning & 648 & 161, 482, 483 & One token on a hot path \\
grok-3 & 474 & 2, 312, 313 & One token on a hot path \\
grok-4 & 1{,}161 & 679, 997, 998 & One token on a hot path \\
grok-4-1-fast-reasoning & 627 & 463, 464, 151 & One token on a hot path \\
grok-4-20-non-reasoning & 640 & 128, 128, 128 & Coarse block, no one token \\
grok-4.3 & 640 & 128, 128, 448 & Coarse block, no one token \\
grok-3-mini & 493 & 2, 329, 331 & Two token step \\
grok-3-mini-fast & 493 & 329, 2, 331 & Gapped \\
grok-4-fast-reasoning & 635 & 471, 151, 146 & Unstable \\
grok-4-20-reasoning & N/A & N/A & Incomplete \\
\bottomrule
\end{tabular*}
\end{table*}

The public prefix corpus was selected by availability of a public harness request prefix.  It contains \texttt{claude\_code}; Cursor variants \texttt{2025-03-09}, \texttt{2025-09-03}, \texttt{agent-cli-2025-08-07}, \texttt{agent2.0}, \texttt{chat}, \texttt{v1.0}, and \texttt{v1.2}; and OpenCode variants \texttt{anthropic}, \texttt{beast}, \texttt{codex}, \texttt{copilot-gpt-5}, \texttt{default}, \texttt{gemini}, \texttt{gpt}, \texttt{kimi}, and \texttt{trinity}.  The frozen Grok-2 tokenizer digest is \texttt{63e538a53de71acf0}; the measurement manifest represents each input by path, byte count, and SHA-256.

A second offline pass over the same corpus measures how much of each prefix an \aone{} probe can reproduce.  The scan counts only values a harness resolves at run time: an expanded date, a home directory path, an operating system version line, a git status block, or an environment block header.  It deliberately excludes static markup tags such as \texttt{<user\_query>} and template prose that merely names a runtime concept, because both are identical on every run; a scan that treats every angle bracket tag as dynamic reports 15 of 17 prefixes as dynamic and inverts the result.  Under the strict rule, 15 prefixes contain no resolved value, \texttt{cursor\_agent-cli-2025-08-07} reproduces 2,952 of 3,019 tokens before an environment block header, and \texttt{cursor\_v1.2} reproduces 7,077 of 7,165 tokens before a resolved home path.  Reproducible spans run from 1,559 to 8,794 tokens with a median of 2,632, and all 17 clear both the 256 and 1,024 token floors on the reproducible span alone.  The pass sends no request and records hashes, offsets, and counts without prompt text.

\begin{table*}[t]
\centering
\caption{Detailed evidence ledger for identity composition.  Table~\ref{tab:main-results} gives the compact synthesis in the main text.}
\label{tab:detailed-identity-ledger}
\footnotesize
\setlength{\tabcolsep}{2.5pt}
\renewcommand{\arraystretch}{1.14}
\begin{tabularx}{\textwidth}{@{}L{0.72}L{0.62}L{1.70}L{0.96}@{}}
\toprule
Unit & Evidence class & Observed outcome & Inference unit \\
\midrule
Six frozen source paths & Audit of source and data flow & Susceptible forwarding recurred across at least five lineages; two projects also contained contrasting adapter paths. & Adapter path under its frozen configuration. \\
Five gateway routing graphs & Audit of source and runtime & All five support multiple downstream principals and a graph with an exclusive credential; ordinary pools use a different selector from the original principal. & Gateway implementation and configuration; provider enforcement is a separate premise. \\
Ten gateway and provider cells & Contract from source to runtime & Each deliberately pooled gateway returned $[0,1408,1408,0]$ through OpenAI and $[0,h,h,0]$ through Anthropic, with $h$ from 1,066 to 1,074. & Five frameworks repeated over two provider families. \\
Principal split & Matched intervention & Two hits under a shared principal became two controlled misses under split principals. & Gateway and principal mapping cell. \\
Provider namespace & Randomized matched intervention & Shared principal and shared namespace crossed 5/5; splitting either variable crossed 0/5, with 15/15 owner hot repeats. & Principal and namespace trial. \\
Recorded credential pool & Selector association & The same recorded partition crossed 3/3; different partitions crossed 0/3. & Cell for the selected partition. \\
Nested relay & Matched intervention & Shared paths and paths with an inner overwrite crossed; preserving the original customer removed reachability across customers.  One real upstream cell per arm; a replay on rebuilt gateways forwarded the predicted namespace and returned the same verdict for five different prompts per arm. & Three nested path arms plus a five prompt replay. \\
Managed/BYOK & Platform bundle intervention & Managed crossed 5/5; independently scoped BYOK crossed 0/5 with 5/5 owner hot repeats. & Account pair and platform arm. \\
OpenRouter ranked existence & Weekly ranking frame selected independently of outcomes plus latest strict protocol & Of 28 eligible labels, 12 have a positive route, one has a negative on the tested path, 11 are inconclusive, and four remain untested; protocol volume coverage is 80.5\%. & Label $\times$ route $\times$ dated window and association with model volume. \\
Relays stratified by framework & Production frame constrained by availability & Four attributed deployments cover two of five strata: three positive, one mixed negative and inconclusive; three strata remain missing. & Service, model, route, account pair, and window. \\
\bottomrule
\end{tabularx}
\end{table*}

The intervention arms carry small valid denominators, so we report exact counts and state what each design supports.  Clopper-Pearson 95\% intervals are $[0.478,1]$ for a 5/5 arm, $[0,0.522]$ for 0/5, $[0.292,1]$ for 3/3, and $[0,0.708]$ for 0/3.

Every arm separated completely, and for a complete separation the exact two sided value depends only on the number of trials: it is 0.008 for five trials against five and 0.10 for three against three, which are the smallest values those sizes allow.  A test statistic therefore adds nothing to the counts unless the design also supplies a reason to treat the compared trials as exchangeable, and only one arm does.  The principal and namespace matrix at Provider~A assigned its trials at random, so comparing either split arm against the shared arm is a randomization test of the hypothesis that assignment does not change the outcome, and we report that pairwise value rather than a joint value over the three arms.  All 15 scheduled trials were valid, so no exclusion conditions the comparison.  The other arms answer different questions and we report them as counts.  Managed and BYOK compare two fixed route classes and move a platform bundle rather than one assigned selector.  The pool arm records the partition the gateway chose instead of assigning it.  The nested replay ran against the local fixture, where repeating a run reproduces the outcome rather than drawing a new one, so it establishes that each gateway forwarded the predicted namespace for five different prompts.

What supports those arms is the matched design: each fixes the prompt, schedule, model, and observer, changes one predicted selector, and retains an owner hot repeat that confirms cache formation, so a predicted direction and a paired control carry the inference.  Repeated estimates under real provider variability would strengthen the pool and nested arms, as Section~\ref{sec:limitations} notes.

\subsection{Production classification and OpenRouter cohort}

We now apply the qualifications above to production services.  Each production cell records coded service, model and provider label, served model and fingerprint when available, account pair, UTC window, protocol family, scheduled and valid trials, strict positive, controlled miss, formation and route outcomes, transport status, negative controls, prompt, completion, and reasoning tokens, errors, retries, and reported versus authoritative cost fields.  Fresh canaries repeat only within a cell.  A transport failure never enters a valid denominator; failed formation remains inconclusive.

The search of coded relays sought at least one independently operated deployment with two accounts and managed credentials per executable project.  Four eligible
deployments were found in two strata.  NewAPI deployment N1 has positive GPT
and Grok paths; N2 has one controlled negative direction
$[0,0,1408,0]$ and one reverse direction inconclusive for formation.  Sub2API deployments S1
and S2 produced 2/9 and 1/8 strict valid hits on their default routes, respectively, with
two further S2 inconclusives.  No eligible uni-api, MetaAPI, or LiteLLM unit was
found, so those three strata remain visible as missing units in the
frame.  Two additional services of unidentified lineage each produced
bidirectional $[0,1024,1024,0]$ positives.  Availability determined retention;
outcomes did not.

The OpenRouter relevance frame freezes daily ranking rows for
July 26 through August 1, 2026 and their union of 28 eligible closed chat model labels.
Only the latest protocol with four requests supplies a result.  Older records from the synthetic gate
are excluded, while duplicate and discordant strict cells remain linked
in the ledger.  Fixed routes, fresh canaries represented by digests, disabled inference retries,
formation verified by an owner hot repeat, and isomorphic fresh negatives define eligibility.

Cells from the latest protocol cover 80.5\% of the frame's listed token
volume.  Twelve labels have at least one positive route; one has a valid
negative on the tested path; eleven are inconclusive because formation, route, or
accounting controls fail; and four were not tested under the latest protocol.
The volume associated with labels that have a positive route is 4.031 trillion tokens, 33.7\% of
volume for eligible models and 7.1\% of OpenRouter's platform denominator.  These
weights describe the popularity of model labels on which a positive path
exists.  Section~\ref{sec:limitations} states the unavailable route, credential, and eligibility denominators.
Across retained strict campaign files, including superseded and corrective
cells, the ledger accounts for 156 requests and \$0.657 in cost reported by responses.

The paired control for length uses three alternating short and long blocks.  Each arm is positive in two blocks, and discordances point in opposite directions.  Exact McNemar $p=1$ summarizes these three pairs.  Positive short probes report 1,792 cached tokens and positive long probes report 3,840, leaving attribution of binary reachability to length or a latent selector unresolved.

\subsection{Defense cells for five gateways}

We next expand the defense matrix used to test separation.  Table~\ref{tab:defense-exact-vectors} expands the compact defense matrix in
Section~\ref{sec:defenses}.  The local fixture canonicalizes bearer and
\texttt{x-api-key} representations, keys its cache by credential or namespace
as the arm requires, and admits no entry on injected failure.  This matrix runs entirely against the local fixture.

The MetaAPI graph shows how easily an exclusive configuration can be written incorrectly.  A pilot policy using a broad \texttt{supportedModels} rule bypassed the intended \texttt{allowedRouteIds} restriction, so we excluded it; the final policy restricted to one route and one channel failed closed as the table records.

\begin{table*}[t]
\centering
\caption{Exact vectors from four requests for frozen local defense graphs.  ``N/I'' marks projects where we added no implementation that derives a namespace on the relay.}
\label{tab:defense-exact-vectors}
\footnotesize
\setlength{\tabcolsep}{3pt}
\renewcommand{\arraystretch}{1.10}
\begin{tabular*}{\textwidth}{@{\extracolsep{\fill}}lcccccc@{}}
\toprule
Gateway & Shared & Honest client & Forged client & Derived by relay & Exclusive & Status after B failure \\
\midrule
NewAPI & $[0,325,325,0]$ & $[0,0,325,0]$ & $[0,325,325,0]$ & $[0,0,325,0]$ & $[0,0,325,0]$ & 200/503/200/503 \\
uni-api & $[0,325,325,0]$ & $[0,0,325,0]$ & $[0,325,325,0]$ & N/I & $[0,0,325,0]$ & 200/503/200/503 \\
MetaAPI & $[0,336,336,0]$ & $[0,0,336,0]$ & $[0,336,336,0]$ & N/I & $[0,0,336,0]$ & 200/503/200/503 \\
LiteLLM & $[0,325,325,0]$ & $[0,325,325,0]$ & $[0,325,325,0]$ & N/I & $[0,0,325,0]$ & 200/503/200/429 \\
Sub2API & $[0,325,325,0]$ & $[0,0,325,0]$ & $[0,325,325,0]$ & N/I & $[0,0,325,0]$ & 200/502/200/502 \\
\bottomrule
\end{tabular*}
\end{table*}

\subsection{Ecosystem denominators}
\label{app:ecosystem-denominators}

We next place the measured paths in the wider relay ecosystem.  Table~\ref{tab:ecosystem-denominators} preserves the distinct units behind the ecosystem relevance argument.  The measurements complement one another and retain their native denominators.

\begin{table*}[t]
\centering
\caption{Independent denominators for relevance of the relay ecosystem.}
\label{tab:ecosystem-denominators}
\small
\setlength{\tabcolsep}{4pt}
\renewcommand{\arraystretch}{1.14}
\begin{tabularx}{\textwidth}{@{}L{0.85}L{1.05}L{1.12}L{0.98}@{}}
\toprule
Frame & Reported quantity & What it measures & Interpretation limit \\
\midrule
OpenRouter status reported by platform & 200T+ tokens/month; 10M+ users & Current platform scale in two native units. & Platform scale with routes and cache state unobserved. \\
External model of global tokens & Approximately 2\% in May 2026 & Estimated OpenRouter share of global monthly LLM tokens. & Modeled platform share for May 2026. \\
Ramp transaction panel & 70,000+ businesses; roughly half of category purchasers used OpenRouter & Purchaser adoption in a large panel of business spending. & Adoption among purchasers in the panel. \\
NIST/CAISI trace & 97.5M requests over four complete weeks~\cite{nist_deepseek} & Requests to one model family through OpenRouter. & One family and observation window. \\
Open source census & 29 retained; 28 unarchived; six audited; five executable & Project ecosystem and frame for selecting projects for depth. & Project availability and adoption proxies. \\
OpenRouter ranked existence & 28 eligible labels; 80.5\% coverage of listed volume; 12 positive routes, one negative on the tested path, 11 inconclusive, four untested & Dated existence over a weekly ranking by token volume selected independently of outcomes. & Popularity of labels and strict existence cells. \\
\bottomrule
\end{tabularx}
\end{table*}

\subsection{Extraction implementation and accounting}
\label{app:extraction-gates}

We now give the implementation and accounting for the extraction experiments.

Section~\ref{sec:methodology} defers the numeric gates to this appendix.  Let $T(q)$ be the reported prompt token count.  A \emph{burner} is the control extension outside the candidate alphabet.  Owner repeats, burner responses, and scored candidate responses all use the reuse threshold $\tau(q)=T(q)-32$, so at most 32 prompt tokens may remain uncached.  The owner formation gate additionally requires the owner repeat to raise $R(q)$ by at least 100 tokens over the prime.  A primary and shadow pair counts as jointly hot only when both responses clear their thresholds and report equal prompt token counts; status, model, route, and retry checks establish the rest of the comparability.  The first read of a shadow context precedes its preheating and therefore reports a cold count by construction, so it records the shadow baseline rather than passing a threshold.

For each verified prefix, the general implementation (1) generates and deduplicates a wide queue of raw strings; (2) increases candidate depth without target token identifiers; (3) merges a structured grammar after detector and judge authorization; (4) constructs primary and no victim shadows with matching tokens; (5) invalidates asymmetric route, formation, usage, status, or retry behavior; and (6) commits under the frozen score or returns \textsc{Abstain}.

We call the frozen live confirmation \emph{Stage C}, after the route discovery and calibration stages that preceded it.  Its $4^8$ specialization fixes candidate orders and target commitment before traffic.  At each position, four gate requests precede 12 candidate requests.  The complete path containing victim state therefore has 32 gate and 96 candidate requests; adding two requests for victim formation and 128 matched requests for replay without a victim yields 258.  The ledger spans roughly 54 seconds and records 295,530 prompt tokens, 258 completion tokens, and zero reasoning tokens.  The relay reported a cost of zero, which we record as response metadata.  Stage C followed successful discovery of the route family and calibration at one and four positions.  Its first eligible execution passed every committed integrity gate and supplies the production existence result.

Table~\ref{tab:stagec-raw-position} exposes all 16 usage reads at the first position.  Every request returned HTTP 200, one completion token, and the same served model.  Candidate delta is primary cached tokens minus the second shadow read; the scorer used only this relative value and the frozen gates.

\begin{table*}[t]
\centering
\caption{Complete usage trace at the first position of the live Stage C chain.  The candidate order was frozen; \emph{bone} was the committed truth.}
\label{tab:stagec-raw-position}
\small
\setlength{\tabcolsep}{4pt}
\renewcommand{\arraystretch}{1.12}
\begin{tabular*}{\textwidth}{@{\extracolsep{\fill}}llrrrrl@{}}
\toprule
Phase/branch & Role & Prompt & Cached & Delta & HTTP & Interpretation \\
\midrule
Frontier & Primary gate & 1,141 & 1,126 & N/A & 200 & frontier containing victim state reachable \\
Frontier & Shadow reference & 1,141 & 168 & N/A & 200 & shadow baseline before preheat, not threshold gated \\
Burner & Primary gate & 1,142 & 1,126 & N/A & 200 & extension outside alphabet adds no lift \\
Burner & Shadow gate & 1,142 & 1,126 & N/A & 200 & burner baseline with matching tokens \\
\midrule
genius & Shadow 1 & 1,142 & 1,126 & N/A & 200 & preheat \\
genius & Shadow 2 & 1,142 & 1,141 & N/A & 200 & reference after preheat \\
genius & Primary & 1,142 & 1,126 & $-15$ & 200 & wrong \\
bone & Shadow 1 & 1,142 & 1,126 & N/A & 200 & preheat \\
bone & Shadow 2 & 1,142 & 1,141 & N/A & 200 & reference after preheat \\
bone & Primary & 1,142 & 1,127 & $-14$ & 200 & unique $\beta+1$; commit \\
picnic & Shadow 1 & 1,142 & 1,126 & N/A & 200 & preheat \\
picnic & Shadow 2 & 1,142 & 1,141 & N/A & 200 & reference after preheat \\
picnic & Primary & 1,142 & 1,126 & $-15$ & 200 & wrong \\
kitchen & Shadow 1 & 1,142 & 1,126 & N/A & 200 & preheat \\
kitchen & Shadow 2 & 1,142 & 1,141 & N/A & 200 & reference after preheat \\
kitchen & Primary & 1,142 & 1,126 & $-15$ & 200 & wrong \\
\bottomrule
\end{tabular*}
\end{table*}

For alphabet size $\alpha$ and $n$ positions, the frozen protocol sends $n(3\alpha+4)$ requests on the path containing victim state and $2+2n(3\alpha+4)$ total controlled requests.  Table~\ref{tab:a2-cost-extrapolation} applies that accounting and the observed mean of 1,145.5 prompt tokens per Stage C request.  A valid BIP-39 schedule with twelve words has 22,656 logical decisions because the preceding entropy and checksum constrain candidates for the last word.  The extrapolation leaves routing success and cache residence as external conditions at larger widths.  Its totals include mirrored attribution and complete replay; an attacker who accepts weaker provenance or a higher error rate can use fewer requests.

\begin{table}[t]
\centering
\caption{Request and token extrapolation from the live integrity protocol.  Attack requests count $n(3\alpha+4)$ requests on the path containing victim state and exclude the two victim formation requests, which the total counts once.}
\label{tab:a2-cost-extrapolation}
\footnotesize
\setlength{\tabcolsep}{2.5pt}
\renewcommand{\arraystretch}{1.12}
\begin{tabular*}{\columnwidth}{@{\extracolsep{\fill}}lrrr@{}}
\toprule
Target & Attack req. & All req. & Input tok. \\
\midrule
$4\times8$ live & 128 & 258 & 0.296M \\
Base62 $\times32$ & 6,080 & 12,162 & 13.9M \\
$2048\times8$ & 49,184 & 98,370 & 112.7M \\
BIP-39, 12 valid words & 68,016 & 136,034 & 155.8M \\
\bottomrule
\end{tabular*}
\end{table}

\subsection{Local candidate and structured studies}
\label{app:local-studies}

The live protocol above isolates sequential composition; local candidate studies now measure the breadth hidden by its four word alphabet.  The corpus with an unknown vocabulary uses schedules of 3,000 probes with the correct previous tokens supplied; revealing the provider vocabulary added about two percentage points of coverage and reduced queries by factors from 1.21 to 1.28 at commonly successful positions.  WildChat selections are unique by source conversation, and completion counts a record as recovered when the procedure reconstructs the full target.  The confirmation with 76 conversations under three tokenizer conditions and fixed endpoints froze records, endpoints, budgets, paired comparisons, and a base seed of \texttt{2026072803}.  At 3,000 probes the known one arm exceeded blind recovery by 45.6 points at endpoint 9, with a conversation cluster 95\% interval from 37.3 to 53.9, and by 39.9 points at endpoint 17, from 32.0 to 47.8.  Intervals clustered by conversation and two sided Monte Carlo sign flip tests use 200,000 flips and a correction of plus one; the reported $5\times10^{-6}$ values are the minimum $1/200001$.  The earlier study with 98 conversations remains separate.

The structured detector frame has 97 positives and 91 hard negatives, and produced 30 true positives, no false positives, and 67 false negatives; a wider judge added two positives and four false positives.  Fixed confusion counts are descriptive and have no population interval.  The experiment over the BIP-39 alphabet uses 16 fixtures with 12 words and invalid checksums, together with 192 positions supplied with the correct previous words.  The scorer receives the public dictionary of 2,048 raw strings but receives neither target token identifiers nor target token lengths.  With reporting at one token, it recovered 192/192 positions and completed 16/16 paths under each of three unknown tokenizer conditions, with zero wrong commits.  At $g=2,4,8$, correct positions fell to ranges from 96 to 97, 48 to 49, and 16 to 17, with no completed path.  The counts at $g=2$ and $g=4$ match the alignment expectation of $192/g$ derived in Appendix~\ref{app:formal-results}, while the count at $g=8$ falls below it because the verified prefix lengths are not uniform modulo eight, so fewer positions land at the reporting boundary.  The three tokenizer conditions repeat the same 16 fixtures.

\subsection{Extraction evidence and eligibility}
\label{app:extraction-scope}

The implementation and local studies now permit a uniform classification of extraction evidence.  Table~\ref{tab:extraction-contract} separates analytic evidence, local components, and live sequential evidence.  No extraction cell carries \evidence{E2}, because a dependent chain needs a production route rather than a configured gateway.  Table~\ref{tab:a2-eligibility} then applies one prospective composition rule to every live or historical extraction record.

\begin{table*}[t]
\centering
\caption{Extraction evidence contract.}
\label{tab:extraction-contract}
\small
\setlength{\tabcolsep}{4pt}
\renewcommand{\arraystretch}{1.14}
\begin{tabularx}{\textwidth}{@{}L{0.35}L{1.15}L{1.45}L{1.05}@{}}
\toprule
Tier & Experiment & Directly supported inference & Independent unit \\
\midrule
\evidence{E0} & Exact attributable extension oracle & Equality search over $\prod_i |C_i|$ becomes at most $\sum_i |C_i|$ logical candidate decisions. & Analytic construction. \\
\evidence{E1} & Simulators over raw strings, replay of a public corpus, structured fixtures, and granularity sweeps & Candidate coverage, effects of an unknown tokenizer, mode switching, local error sensitivity, and conditional identifiability. & Document, conversation, fixture, or synthetic path as declared. \\
\evidence{E3} & Committed relay chain through a model with closed weights and complete matched replay without victim state & Eight dependent decisions made without access to the answer composed in one production $4^8$ cell satisfying every integrity gate. & One complete chain. \\
\bottomrule
\end{tabularx}
\end{table*}

\begin{table*}[t]
\centering
\caption{Prospective eligibility of production extraction records.  A formal chain requires dependent decisions made without access to the answer, comparable primary and shadow histories, complete replay, and successful integrity gates.}
\label{tab:a2-eligibility}
\small
\setlength{\tabcolsep}{3pt}
\renewcommand{\arraystretch}{1.12}
\begin{tabularx}{\textwidth}{@{}L{0.90}L{0.95}L{0.60}L{1.90}C{0.65}@{}}
\toprule
Record & Prefix evolution & Matched replay & Outcome & Formal chain \\
\midrule
Stage C $4^8$ & Dependent commits without access to answer & Complete & Eight commits; zero abstentions or wrong commits & Yes \\
Strict runs with eight candidates & Hidden answer until integrity failure & Incomplete & Truth retained the unique lift at 7/7 positions; both executions abstained after responses fell below the reuse threshold & No \\
Diagnostic with position reset & Correct prefix externally restored & Incomplete & Prospective forced scorer correct 3/6; filter defined after the run accepted 2/6, both correct & No \\
Legacy $P(6,3)=120$ & Older controls & Incompatible & Three correct, two clean stops, zero wrong commits & No \\
Records before May 15 & One position or legacy programs & Incompatible & Later provenance audit retained 0/3 old successes & No \\
\bottomrule
\end{tabularx}
\end{table*}

\subsection{Abstaining and historical extraction evidence}

The prospective eligibility rule above determines how abstaining and historical records contribute to the claim.  Two executions with eight candidates evaluated seven positions, so each execution read 56 primary candidate responses.  The truth retained the expected unique lift in all seven, but one wrong candidate response was cold in each execution; both executions abstained under the rule that requires every response to be hot.  The records localize route integrity as the failed condition while retaining the signal dependent on the candidate.  A prospectively frozen run that tolerated erasures failed its public route gate before candidate reads.  In a later diagnostic that restored the correct prefix at every position, prospective forced selection was correct in 3/6 positions, while a filter defined after the experiment accepted 2/6 and both accepted values were correct.  These diagnostics remain outside the formal chain.

Records before May 15 contain three success verdicts from an older program on synthetic subsets with one position and other signals for ranking candidates.  A later stateful provenance audit found attacker cache writes ambiguous in every comparable primary run, so 0/3 old successes satisfy the present standard for attribution to victim state.  The older $P(6,3)=120$ experiment, with three correct decisions, two clean stops, and zero wrong decisions, remains historical bounded evidence because its controls differ from the complete $4^8$ replay.  The current formal denominator excludes all of these records.

\subsection{Timing and reproducibility}

We conclude with the remaining timing result and reproducibility materials.  The retrospective timing set contains 380 labeled observations without streaming, comprising 151 hits and 229 misses.  The area under the receiver operating characteristic curve (AUC), made direction free and centered within each run, is 0.518; balanced accuracy when leaving out one run at a time is 0.564.  The dataset lacks randomized streaming measurements of time to first token, contemporaneous paired misses, and tail calibration, so timing supports no core attack claim.

Canonical analyses emit results readable by machines.  The claim ledger marks supported, conditional, inconclusive, and withdrawn claims. Historical fingerprint probes that could populate their own cache, inference about traffic from real users, language about functional credentials, and broad priority claims are tombstoned.

\end{document}